\documentclass[aps,prb,twocolumn,superscriptaddress,showpacs]{revtex4}
\usepackage[dvips]{graphicx}
\usepackage{dcolumn}
\usepackage{bm}
\newcommand{\beqn}{\begin{equation}}
\newcommand{\eeqn}{\end{equation}}
\newcommand{\bet}{\begin{tabular}}
\newcommand{\ent}{\end{tabular}}

\begin{document}
\title{Statistical analysis of Ni nanowires breaking processes : a  numerical simulation study}

\author{P.\ Garc\'{\i}a-Mochales}
\affiliation{Departamento de F\'{\i}sica de la Materia Condensada, Facultad de Ciencias, Universidad Aut\'onoma
de Madrid, c/ Francisco Tom\'as y Valiente 7, Campus de Cantoblanco, E-28049-Madrid, Spain}
\email{pedro.garciamochales@uam.es}
\date{\today}
\author{R.\ Paredes} 
\affiliation{Centro de F\'{\i}sica, Instituto Venezolano de Investigaciones Cient\'{\i}ficas, Apdo 20632, Caracas 1020A, Venezuela}
\affiliation{Instituto de Ciencia de Materiales de Madrid,\\
Consejo Superior de Investigaciones Cient\'{\i}ficas\\
c/ Sor Juana In\'es de la Cruz 3, Campus de Cantoblanco, E-28049-Madrid, Spain}
\author{S. Pel\'aez} 
\author{P.\ A.\ Serena}
\affiliation{Instituto de Ciencia de Materiales de Madrid,\\
Consejo Superior de Investigaciones Cient\'{\i}ficas\\
c/ Sor Juana In\'es de la Cruz 3, Campus de Cantoblanco, E-28049-Madrid, Spain}
\date{\today}

\begin{abstract}
We have performed the statistical analysis of the breaking behavior of Ni nanowires.
Using molecular dynamic simulations based on the embedded atom method approximation, we have simulated thousands of nanocontact breakages, determining the time evolution of both the nanowire atomic structure and the nanowire minimum cross section time evolution, $S_m(t)$. The accumulation of thousands of breaking events allows the construction of $S_m$ histograms. These histograms allow the analysis of the influence of the temperature, the crystalline stretching direction and the initial nanowire size on the breaking processes. We have also calculated the proportion of monomers, dimers
and more complex structures at the latest stages of the breaking process. We have found important differences among results obtained for different nanowire orientations and sizes, illustrating that very small wires do not capture the behavior of actual thick nanonecks obtained in experiments. Three main cases have been observed: 
(a) For the [111] stretching direction and large nanowire sizes, the wire gradually evolves from more complex structures to monomers and, in several cases, to dimers prior its rupture. This case presents well ordered structures during the breaking process. 
(b) On the other hand, for large nanowires stretched along the [100] and [110] directions the system mainly breaks directly from complex structures, with a low probability of finding monomers and dimers. In this case a clear dependence on temperature is observed. At room temperature a huge histogram peak around $S_m=5$ appears, showing for first time in the literature the presence of long staggered pentagonal Ni wires with ...1-5-1-5-1-5-.... structure. Additionally, we have found for this case that nanowires have disordered regions during their breakage.
(c) When the initial wire size is small strong size effects are observed independent on the temperature and stretching direction used. 
Finally, we have found that the local structure around monomers and dimmers do not depend on the stretching direction and that these configurations differ from those usually chosen in static studies of conductance.
\end{abstract}


\maketitle


\section{Introduction}

Nanotechnology involves the design, synthesis, fabrication, characterization, and application of structures  by controlling their composition, shape and size at the nanometer scale. The control of these properties will allow to exploit a whole set of novel physical and chemical features in future devices with a wide range of applications. In particular,  electron transport through future nanofabricated devices is a crucial issue, and different candidates have been proposed as potential nanoelectronics elements with conducting or semiconducting properties. Among these candidates, metallic nanowires and nanocontacts are promising disipationless nanodevices as well as exigent benchmarks to test nanoscale basic phenomena \cite{SerenaBook97}. 

The interest on metallic nanowires and nanocontacts rises from their very rich phenomenology.
Electron transport through metallic nanowires present ballistic features at and below room temperature (RT) since, in general, the electron inelastic mean free path is larger than the characteristic nanowire dimensions.
Furthermore, well defined electron transport modes or channels appear associated to the transversal confinement of electrons for nanowires with diameters of the order of few Fermi wavelengths $\lambda_F$.
In fact, a narrow metallic nanowire is a type of mesoscopic system that is characterized by the existence of many electron levels (whose exact distribution depends on the precise atomic configuration) coupled to the electron energy spectra of the reservoirs. In the ballistic limit, the conductance $G$ (inverse of the nanowire resistance $R$, $G=1/R$) is described in terms of the Landauer formula\cite{LandauerPhilMag70}:

\begin{equation}
\label{landauer1}
G=(G_0 / 2)\sum_{\sigma} \sum_{n=1}^{N_{\sigma}} T_{n {\sigma}}
\end{equation}

\noindent
where $G_0$=2e$^2$/h is the conductance quantum (e being the electron charge and h the Planck constant), ${\sigma}$ denotes the two possible spin polarizations, $N_{\sigma}$ is the spin-dependent number of accessible transport channels with energies below the Fermi energy, $E_F$, and $T_{n \sigma}$ is the spin-dependent transmission probability associated to the $n$-th mode. The calculation of the quantities $T_{n {\sigma}}$ is a difficult task since the solution to the scattering problem requires the full description of nanocontact band structure as well as the knowledge of those scattering mechanisms (geometry changes, defects, magnetic domain walls, etc) governing the electron transport. Consequently, the electron transport is governed by the set of the non negligible transmission probabilities associated to the propagating modes. Notice that for degenerate energy levels Eq. (\ref{landauer1}) is written as $G=G_0 \sum_{n=1}^{N} T_{n}$, where spin dependences disappear.

Several experimental techniques have been used to form metallic nanocontacts and nanowires\cite{SerenaBook97}. Scanning tunneling microscopy (STM)\cite{PascualPRL93,AgraitPRB93,OlesenPRL94,OlesenPRL95} and mechanically controllable break junction (MCBJ)\cite{MullerPRL92,KransNature95} methods are standard  approaches to study, in a rather controlled way, the process of formation and rupture of nanocontacts under different experimental conditions. Metallic nanowires are also obtained using electron-beam irradiation inside ultra high vacuum (UHV) transmission electron microscopes (TEM)\cite{KondoPRL97}. 
This procedure allows the direct monitoring of the metallic nanowires thinning process upon external perturbations (stretching, heating, etc). In some cases, this technique has revealed that nanowires with few atoms in the cross section present helical and weird structures\cite{KondoSci00}. Metallic nanowires are also formed through electrochemical methods by controlling the potential between pairs of electrodes immersed in electrolytic solutions\cite{LiAPL98,ElhoussineAPL02}. Finally, methods based on 'table--top' experiments (where contacts are broken by separating two macroscopic metallic wires)\cite{CostaSS95,GillinghamJPCM02} provide a simple but uncontrolled way to generate metallic nanowires.

The study of both ballistic and quantum features associated to the electronic transport through metallic nanowires requires their full electric characterization. This is usually done during the formation and/or the rupture of a metallic nanowire, allowing the access to many atomic configurations within the same experiment, and obtaining the nanowire conductance time evolution during its formation (or breaking) process, i.e. the so-called conductance trace, $G(t)$.  Each conductance trace presents its own features since it is very difficult to accurately control the nanowire geometry during its mechanical deformation. 

In order to obtain relevant information concerning the electronic transport through nanowires of a given metallic species, conductance histograms\cite{OlesenPRL95} $H(G)$ are constructed by accumulating hundreds of conductance traces obtained under the same experimental conditions. 
Conductance histograms $H(G)$ provide valuable statistical information on the transport and structural nanocontacts properties. In many cases, conductance histograms present well defined peaked structures close to integer multiples of $G_0$ , reflecting the existence of preferred conductance values. Such preferential values are usually interpreted in terms of conductance quantization\cite{YansonNature99} or  favorable atomic arrangements\cite{YansonPRL97,HasmyPRL01,DiazNanotech01}. 

The interpretation of conductance histograms is a difficult task since they merge mechanical and electrical information. Moreover, both properties are strongly correlated\cite{RubioPRL96}. In fact, those atomic rearrangements taking place during the nanowire stretching process modify the number of accessible modes as well as the set of transmission coefficients, leading to conductance changes. The interpretation of conductance histograms is even more intrincated for polyvalent metals since several channels per atom are involved in the electronic transport\cite{ScheerPRL97,ScheerNature98}. For instance, aluminum conductance histograms obtained at both 4K\cite{YansonPRL97} and RT\cite{DiazNanotech01} show well defined peaks at conductance values close to integer values of $G_0$. However, the first conductance histogram peak appearing at $G \sim G_0$ presents contributions from three propagating channels\cite{ScheerPRL97}.

In spite of these interpretation problems, conductance histograms have become a standard tool to analyze the dependence of the nanowires electron transport on several experimental parameters as temperature\cite{YansonNature99,YansonLTP01}, applied voltage\cite{YasudaPRB97}, adsorbed species\cite{LiPRB98}, and applied electrochemical potential\cite{ShuPRL00}. Furthermore, conductance histograms have been used to detect magic structures for different metallic nanowires\cite{YansonNature99,YansonLTP01,YansonPRL01,MedinaPRL03,MaresPRB04}.

The simultaneous control of charge and spin currents using a suitable combination of geometries, applied magnetic fields and bias voltages opens an interesting landscape for future applications. 
For this reason, electronic transport in atomic sized magnetic nanowires has been profusely studied\cite{OlesenPRL94,OlesenPRL95,SirventPRB96,CostaPRB97,OttPRB98,OshimaAPL98,OnoAPL99,KomoriJPSJ99,GarciaPRL99,TataraPRL99,LudophPRB00,CostaRMF01,OokaJMMM01,ShimizuJMMM02,ElhoussineAPL02,ChopraPRB02,ViretPRB02,BakkerPRB02,RodriguesPRL03,ElhoussineJAP03,UntiedtPRB04,SekiguchiJMMM04,BrandbygePRB95,SekiguchiJAP05,DiazJMMM06,CalvoIEEE07,UntiedtPRL07}.  However, the analysis of conductance histograms $H(G)$ becomes even more intricate in magnetic nanowires  due to the presence of this new degree of freedom as well as the presence of new scattering sources as magnetization domain walls.

There are substantive studies on nickel nanowires, in comparison to those referred to cobalt or iron, for instance. In particular, several experimental  works have addressed the study of nickel conductance histograms under different environmental conditions. Although for most metals (notably Au) the histograms reported by different groups are very similar, there exists a profusion of different $H(G)$ curves for Ni. Therefore, it seems that Ni conductance histograms strongly dependen on the experimental conditions as well as on the nanowire fabrication methodology.

The first study where a conductance histogram was reported\cite{OlesenPRL95} also claimed the existence, at RT, of a Ni conductance histogram very similar to that of platinum, i.e. showing a peaked structure. These histograms were constructed using few conductance traces and applying a selective criterion to choose them. However, further experiments (performed at RT, without applied magnetic field, and by adding thousands of unselected conductance traces) obtained featureless Ni conductance histograms\cite{CostaPRB97,OttPRB98,CostaRMF01}. 

Very different results, showing well defined peaks, were reported for Ni conductance histograms\cite{OshimaAPL98} obtained in UHV, at room and higher temperature, and constructed from many conductance traces. These histograms showed evidences of fractional conductance quantization as expected for magnetic systems characterized by ballistic conductance and the presence on non-degenerated states (see Eq.\ \ref{landauer1}). More evidences of fractional conductance were obtained in several experiments carried out at RT where histograms were built up from few selected conductance traces.\cite{OnoAPL99,OokaJMMM01,ShimizuJMMM02,SekiguchiJAP05} Histogram peaks located at noninteger values of $G_0$ have also been observed during the formation of Ni nanowires by electrodeposition\cite{ElhoussineAPL02}. Similar results have been reported for other magnetic metallic species\cite{KomoriJPSJ99,RodriguesPRL03} and for nanowires made of bulk non-magnetic metals\cite{RodriguesPRL03,GillinghamJPCM02} metallic nanowires. The later results could also point toward the presence of spontaneous magnetization in the nanowire\cite{ZabalaPRL98}.  However, it has been recently demonstrated that the use of sets formed by few selected conductance traces allows the construction of histograms with fractional and non-fractional (i.e integer) conductance peaks for Ni and Cu\cite{DiazJMMM06}. Therefore, conductance histograms formed with few conductance traces must be cautiously considered.

A different set of experiments, carried out at 4K, in UHV conditions, and using thousand of conductance traces, showed  Ni conductance histograms with two well defined peaks around $\sim 1.6 G_0$ and $\sim 3.1 G_0$, respectively.\cite{BakkerPRB02,UntiedtPRB04,CalvoIEEE07}.  The peaks were not modified by the presence of strong magnetic fields\cite{UntiedtPRB04}. The first peak position is consistent with pioneering jump-to-tunnel experimental results, obtained at T=4.2K\cite{SirventPRB96}. More recent experiments\cite{CalvoIEEE07} show that the first peak is, in fact, formed by the superposition of two sub-peaks located at $G \sim 1.2 G_0$ and $ G\sim 1.5 G_0$. 

It is clear that there is a broad landscape with different Ni conductance histograms depending on the temperature, the applied voltage, environmental conditions, etc. A similar situation happens for iron nanocontacts\cite{OttPRB98,KomoriJPSJ99,LudophPRB00}. In summary, we can describe two opposite results for Ni when histograms are built without selecting conductance traces: on one hand the low temperature conductance histograms showing a well defined and reproducible peaked structure (not modified by the presence or absence of applied magnetic fields), and, on the other hand, the room temperature conductance histograms, showing a bunch of different experimental results. The difference between low temperature (4K) and RT Ni conductance histograms is not well understood since, in both cases, the system is below its bulk Curie temperature ($T_c$=903K) and a similar magnetic behavior is expected. 

The previously sketched experimental landscape contrasts with more defined theoretical results. An infinite monoatomic Ni nanowire is characterized by a high conductance $G=3.5 G_0$\cite{SmogunovSS02}. This value decreases to $G \sim G_0$ in presence of magnetic walls due to the $d$-states blocking effect\cite{SmogunovSS02}. The same blocking effect is reported for more realistic Ni nanowires formed by three atoms attached to two (100) leads. In this case the conductance takes values comprised in the range $1.35 G_0 - 1.6 G_0$\cite{SmogunovPRB06}. It has been also reported that the conductance takes values in the range $G_0$--$2.5G_0$ when different monomer and dimer configurations are proposed to describe the nanocontact region\cite{SirventPRB96}. More recent calculations on dimer-like Ni nanocontacts have demonstrated  that $G\sim 1.8 G_0$ for a full ferromagnetic situation whereas  $G \sim 1.4 G_0$ in presence of a domain wall\cite{JacobPRB05}. Therefore there are many theoretical results that could explain the first experimental peak appearing in Ni histograms at low temperature. However, the second peak of the conductance histograms remains unexplained.

Encouraged by the diverse experimental results, we have focused the present study on the role played by the mechanical behavior during the breaking process of Ni nanowires at low and room temperatures. Both temperatures are well below the Ni melting temperature ($T_m = 1728$K) and, in principle, a similar mechanical behavior is expected. However, we need to exclude this thermal effect as the origin 
of those marked differences between low and room temperature $H(G)$.  The aim of the present work is to carry out a statistical study of the structural evolution of Ni nanowires under stretching at low and room temperatures following a well tested methodology.\cite{HasmyPRL01,DiazNanotech01,MedinaPRL03,HasmyPRB05,GarciaMochalesAPA05}. We will address the  construction of minimum cross-section histograms from hundreds of computational traces, taking into account different crystallographic orientations and nanowire sizes. We have not limited ourselves to the study of computational minimum cross section histograms but we have also analyzed the different types of structures appearing at the last stages of the breaking processes. An additional motivation for carrying out the present study is also related with the disagreement among different computational histograms that have been recently reported for Ni nanocontacts\cite{GarciaMochalesAPA05,PaulyPRB06,CalvoIEEE07}. These discrepancies must be understood in order to validate the computational construction of histograms as an useful methodology.

The manuscript is organized as follows: in the following section we describe the computational approach used to construct computational histograms.  In Section III, we present the dependence of such histograms on several parameters (temperature, nanowire stretching direction, and nanowire size), paying attention to the different types of atomic configurations appearing at the last stages of the breaking process. Finally, the main conclusions will be described in Section IV.

\section{Computational Methods}

The study of those preferred configurations appearing in nanometric systems as well as their evolution under modification of external parameters is of capital interest. An accurate modeling of nanoscale systems will allow huge savings in expensive fabrication and characterization procedures. The main goal of this study is to understand the reation between the nanowire geometry time evolution and the structure of the conductance histograms. We follow a similar strategy to the experimental one, simulating hundreds of independent breaking events, in order to determine the presence of preferred configurations leading to peaked structures in the conductance histogram. At this point, we should remember that the quantity $H(G)\Delta G$ (where $\Delta G$ is the histogram bin width) is proportional to the time that a given conductance value $G$ is experimentally observed, and, therefore, a conductance peak corresponds to a relatively stable conductance associated to one or several stable atomic configurations. 

The simulation of metallic nanowire breaking events has been carried out using standard Molecular Dynamics (MD) methods based on a semi-classical description of the atom--atom interaction. MD has been extensively used to study the structure and rupture of metallic nanowires. In fact, pioneering MD simulations\cite{LandmanSci90,SuttonJPCM90} were used in the past to predict the abrupt changes in forces arising in a tip-surface indentation/stretching process. 

Ab-initio based methods have been used for describing the dynamics of a single metallic nanowire under stretching\cite{LandmanZPD97}. In spite of their accuracy, the large computational resources required by ab-initio based methods make them unappropriated to determine computational histograms. A high computational effort is also required when other approaches are used as tight-binding molecular dynamics (TB-MD) simulations\cite{daSilvaPRL01}.  It is also possible to determine computational conductance histograms taking into consideration simple models.\cite{TorresPRL96} In general, these methods do not account for a realistic treatment of the nanowire dynamics under stretching. Therefore, in order to perform the statistical analysis of many breaking events, semi-classical MD methods represent a midterm between accuracy and existing computational capabilities. 

In the past years, MD simulations dealing with metallic nanowires mainly focused on the description of single formation or breaking events, using different interatomic potentials but neglecting the study of statistical effects\cite{OlesenPRL94,LandmanSci90,BratkovskyPRB95,SorensenPRB98,IkedaPRL99,BranicioPRB00,BahnPRL01,HeemskerkPRB03,WangPhysE05,ParkPRL05}. 
More recently, several MD studies on breaking and formation processes of nanocontacts have statistically analyzed  the appearance of preferred atomic configurations in order to seek correlations with peaks noticed in experimental conductance histograms at different temperatures.\cite{HasmyPRL01,DiazNanotech01,MedinaPRL03,HasmyPRB05,CalvoIEEE07,GarciaMochalesAPA05,DreherPRB05,PaulyPRB06} Some of these MD simulation studies\cite{HasmyPRB05,DreherPRB05,PaulyPRB06} have been done using an hybrid scheme, where MD configurations are used as starting point for the calculation of the electronic transport using accurate methods. These studies reveal that there exists a complex relation between the atomic configuration and the conductance histograms. For instance, for Al nanocontacts\cite{HasmyPRB05} the first conductance peak around $G \sim G_0$ is caused by dimer-like configurations whereas the second peak close $G \sim 2G_0$ is due to monomer-like configurations. 

We have simulated the nanowire dynamics using a MD scheme where atomic interaction is represented by
embedded atom method (EAM) potentials\cite{DawPRL83}. In EAM, the potential energy function for the system reads:
\begin{equation}
E=\frac{1}{2}\sum_{ij; i \ne j}\phi(r_{ij})+\sum_i F(\bar\rho_i)
\label{EAM}
\end{equation}

\noindent
where {\it i} and {\it j} run over the number of atoms. In the first term, $\phi(r)$ corresponds to a pair potential depending only on the distance $r_{ij}$ between every pair of atoms {\it i} and {\it j}. The second term is the so-called {\it embedding energy}, which depends on the mean electronic density $\bar\rho_i$ at atom {\it i}'s location. This density is approximated in EAM as the sum of the contributions due to the surrounding atoms, $\bar\rho_i=\sum_{j\neq i}\rho(r_{ij})$. Then, the embedding energy is calculated by evaluating and summing the {\it embedding function} $F(\rho)$ at each atom's position. Depending on the material and the specific physical properties to be studied, different pair potential $\phi(r_{ij})$, embedding $F(\rho)$ and density $\rho(r_{ij})$ functions can be defined.
In the present study, we have used the EAM parameterization proposed by Mishin {\it et al.}\cite{MishinPRB99}, which is constructed by fitting almost 30 experimental and ab-initio values corresponding to several bulk and surface properties. We have already used this EAM approach in previous works to describe static and dynamical properties of Al and Ni nanowires.\cite{HasmyPRL01,DiazNanotech01,MedinaPRL03,HasmyPRB05,GarciaMochalesAPA05,PelaezPSS06} Although the simulation does not include explicitly the possible magnetization of Ni, its magnetic nature has been taken into account through the EAM potential since the embedding energy was fitted to some quantities obtained from spin-polarized ab-initio calculations\cite{MishinPRB99}. Furthermore, such magnetic effects are negligible on the nanowire dynamics due to the weakness of magnetic forces with respect to the crystalline field.

Nanowire dynamics have been studied at constant temperature $T$ using a standard velocities scaling algorithm at every MD step. Two temperatures (4 K and 300 K) have been considered in this work, in order to describe the Ni nanowire dynamics at low and room temperatures. The MD time step is $\delta t=10^{-2}\ ps$. Atomic trajectories and velocities were determined using conventional Verlet velocity integration algorithms. The simulation of a single nanowire breaking event consists of three stages: i) the definition of a bulk-like unrelaxed nanowire; ii) a MD relaxation stage of the initial structure; and iii) a MD stretching process at constant velocity till the nanowire breaks.

\begin{table}[tb]
\begin{ruledtabular}
\begin{tabular}{|ccc|ccc|ccc|}
 & [111]& &  & [100] & & & [110] &  \\
\hline
$N_s$ & $N_L$ & $N_{a}$ & $N_s$ & $N_L$ & $N_{a}$ & $N_s$ & $N_L$ & $N_{a}$ \\
\hline
 56 & 18 & 1008 & 49 & 21 & 1029 & 35 & 29  & 1015 \\ 
 30 & 13 &  390 & 25 & 25 &  375 & 20 & 21  &  420 \\ 
 16 & 10 &  160 & 16 & 12 &  192 & 12 & 17  &  204 \\
  9 & 10 &  90  &  9 & 10 &   90 &  9 & 14  &  126 \\
  6 & 10 &  60  &  6 & 10 &   60 &  6 & 14  &   84 \\
\end{tabular}
\end{ruledtabular}
\caption {\label{table1} Geometric parameters of the initial non-stretched nanowires with parallelepiped shape. The total number of atoms in the nanowire ($N_{a}$) (including the frozen slabs) is calculated by multiplying the number of layers ($N_L$) which are perpendicular to the stretching direction [ijk] by the number of atoms per layer ($N_s$).}
\end{table}

The first stage corresponds to the definition of the initial unrelaxed structure. We consider a bulk super-cell with  parallelepiped shape that contains hundreds of atoms ordered according to a $fcc$ structure with bulk Ni lattice parameter $a$=3.52 $\AA$. The initial parallelepiped height ($L$) will correspond to the stretching direction and is larger than the base edges. We define the $z$ axis as the stretching (pulling) direction. In the present study we have considered three different stretching directions corresponding to the [111], [100] and [110] crystalline directions.  In addition, five different parallelepiped sizes have been considered for each stretching direction. Table \ref{table1} shows the geometric parameters characterizing the different initial bulk-like nanowires we have taken into account. For a given stretching orientation the three largest initial configurations have equal aspect ratio. The two smallest configurations have been chosen in such a way that all the configurations present similar initial cross section sizes independently on the stretching direction. The use of three different crystallographic stretching directions as well as five different initial cross-section sizes will allow to study the influence of these factors on the nanowire breaking mechanics at the two chosen temperatures. At the beginning of the simulation the velocity of each atom is assigned at random according to the Maxwellian distribution that correspond to the simulation temperature.

The second stage corresponds to the relaxation of the bulk-like initial structure. Firstly, we define two supporting bilayers at the top and bottom of the supercell. Notice that data shown in table \ref{table1} includes the supporting bilayers atoms. Atomic $x$ and $y$ coordinates within these bilayers will be kept frozen during the simulation. The nanowire will remain attached to these two bulk-like supporting bilayers during the relaxation stage. This stage lasts for 3000 MD steps in order to optimize the nanowire geometry.  

During the third stage, the $z$ coordinate of those atoms forming the top (down) frozen bilayer are forced to increase (decrease) a quantity $\Delta L_z = 10^{-4}\ \mathring{A}$ after every MD step. This incremental process simulates the separation of the supporting bilayers in opposite directions (symmetric stretching) at a constant velocity of $2\ m/s$, giving rise to the subsequent nanowire fracture.
Those atoms non located at the supporting slabs move following the EAM forces. Notice that the stretching velocity is much larger than that used in experiments. This drawback is common for all the computational MD approaches (ab-initio or semiclassical). However, our computational description of the nanowire breaking event is comparable to the "slow" dynamics of actual experimental traces since the stretching velocity is smaller than the sound speed in nickel, allowing a nanowire evolution that follows a set of metastable equilibrium configurations. 

During the stretching stage, the accurate knowledge of the atomic coordinates and velocities allows the determination of many physical quantities characterizing the geometry and energetics of the evolving nanocontact. In particular, we have payed attention to the minimum cross section $S_{m}$ since it is closely related to the favorable configurations appearing during the nanowire evolution under stretching. Furthermore, $S_m$ is closely related to the maximum number of propagating channels and it determines a first--order approximation of the conductance $G$ using semiclassical approaches\cite{SharvinSPJEPT65}. 

The minimum cross section $S_{m}$ is calculated in units of atoms following standard procedures that have been successfully used in previous studies\cite{BratkovskyPRB95,SorensenPRB98}. In our case, we define the radius $r_{0}$ to be equal to half the $fcc$ (111) interplanar distance ($r_{0}=d_{111}/2$). We assign a volume $ V_0 =4 \pi r_{0}^{3} / 3$ to each atom.  To calculate the section at a given $z_{i}$ position, we compute the total atomic volume $V_{tot,i}$ inside a "detecting slab or cursor" with width $\Delta z$. We have used $\Delta z=d_{111}$. Notice that this cursor definition has been also used in previous studies on nanowires stretched along different crystallographic directions\cite{DreherPRB05,PaulyPRB06}. In the present study we keep the same cursor size $\Delta z$ for all the nanowire crystalline directions. This allows a true comparison between histograms obtained for different orientations. The quantity $ S_i = V_{tot,i} / V_0$ corresponds to the section (in number of atoms) at the $z_i$ position.  The detecting cursor moves along the $z$-axis between the two frozen bilayers, using a step equal to $0.1 \times d_{111}$. This scanning-like process determines the cross section $S_i$ as a function of $z_i$ for a given computational time. Finally, from the set of collected $S_i$ values we determine the minimum cross section $S_{m}$ at that time.   In our study $S_m$ is calculated every 10 MD steps. We consider that the nanowire breaking process is completed when $S_{m}=0$. 

\begin{figure}[tb]
\includegraphics{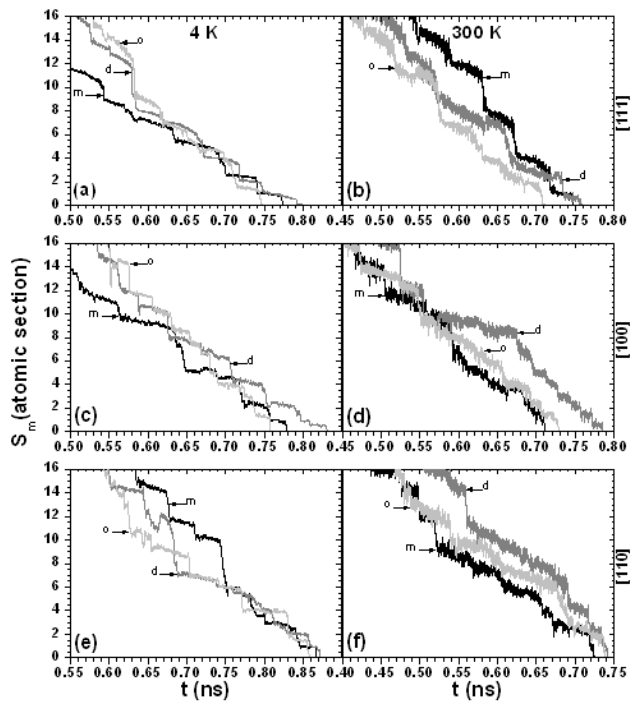}
\caption {\label{fig1} Evolution of the minimum cross section $S_{m}$ for Ni nanowires stretched 
along the [111] (a-b), [100](c-d) and [110] directions (e-f), at 4
K (a, c, e) and 300 K (b, d, f). This figure shows traces corresponding to the largest initial
configurations for each direction (1008, 1029 and 1015 atoms for the [111], [100] and [110]
orentations, respectively). These $S_{m} (t)$ curves illustrate different breaking behaviours. Labels 'm', 'd' and 'o' refer to monomer, dimer, and other structures; see text and Fig.\ \ref{fig2}.}
\end{figure}

The evolution of $S_{m}$ with time shows a typical staircase trace as those examples shown in Fig.\ \ref{fig1}. These curves correspond to the largest simulated nanowires and they have been selected to illustrate different configurations appearing just before the nanowire breaking point. $S_m (t)$ traces show a stepped profile, with well marked jumps associated to atomic rearrangements of the nanowire. We have verified that these jumps are correlated with jumps in the force acting on the supporting slabs. In general, $S_m$ monotonically decreases between two subsequent jumps reflecting the existence of elastic stages. These elastic stages have been associated to the experimentally observed conductance plateaus. $S_{m}$ features larger fluctuations at room temperature than at 4K for the three studied nanowire orientations.

\begin{figure}[tb]
\includegraphics{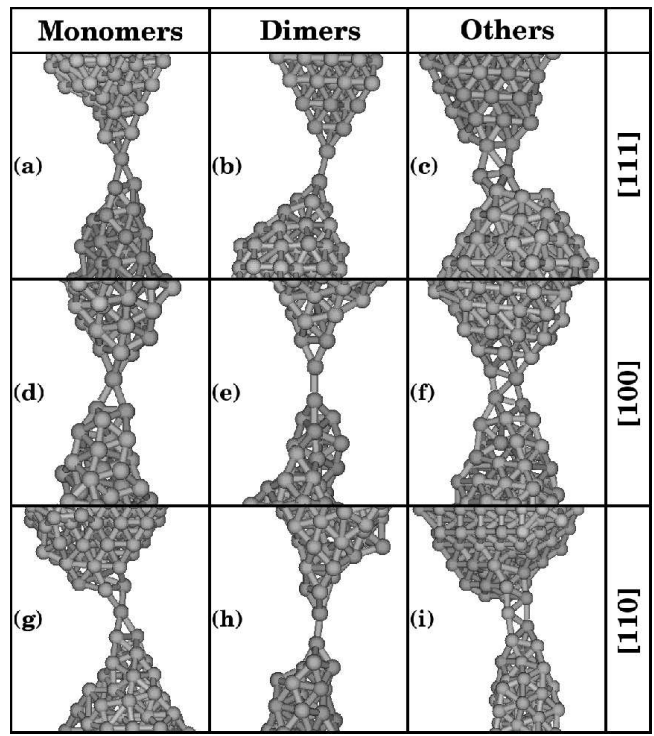}
\caption {\label{fig2} Nanowire configurations just before breakage at T=4K and for three crystalline nanowire orientations  [111], [100] and [110]. Structures were taken from those $S_{m}(t)$ traces shown in Fig. \ref{fig1}. Images illustrating the [111] orientation (a,b,c) correspond to times $t=0.77$, $0.79$, $0.74$ $ns$ for monomer, dimer and other configurations cases, respectively. For the [100] direction, snapshots (d,e,f) correspond to times $t=0.77$, $0.83$, $0.75$ $ns$, respectively. For the [110] stretching direction, snapshots (g,h,i) were taken at $ t=0.86$, $0.87$, $0.85$ $ns$, respectively. }
\end{figure}

The present computational method has been previously tested to simulate nanowire ruptures \cite{HasmyPRL01,DiazNanotech01,MedinaPRL03,HasmyPRB05,GarciaMochalesAPA05}. However, this method poses some problems when used to detect the presence of atomic chains, dimers and monomers, since all these configuration are characterized by $S_{m}$ values close to $\sim 1$. Moreover, the correlation between $S_m$ and conductance is unclear for low $S_m$ values\cite{DreherPRB05,HasmyPRB05,PaulyPRB06}. Therefore, in the present study we have focused our attention on the determination of $S_m$ as well as on the type of atomic configurations formed during the final steps of the breaking stage. We have found that final stages can be classified according to three different categories: monomers, dimers and "others".  These categories are illustrated in Fig.\ \ref{fig2}, where we have depicted the narrowest part of the nanocontact before its breakage for those curves (obtained at T=4K) shown in Fig.\ \ref{fig1}. 

The monomer structure is characterized by a central atom standing between two "pyramids". This configuration usually shows a plateau around $S_{m} \sim 1$. In the dimer structure, the apex atoms of two opposite pyramidal configurations form a two atom chain. In some cases we have found that the dimer is not parallel to the stretching direction. This behavior is caused by the relative shift between the two opposite pyramids during the nanocontact evolution. Depending on the relative angle formed between the dimer and the stretching direction we find values of $S_{m}$ ranging from $\sim 0.5$  to $\sim 1.0$. Final configurations, presenting more complex structures, that do not match these two categories
have been labeled as "others". In this cases we generally observe an abrupt jump from $S_{m} \geq 2$ (i.e., structures formed by two or more atoms) to $S_m =0$. As expected\cite{BahnPRL01}, we have not found any atomic chain formed by three or more atoms for the three studied stretching directions.

To detect the presence of monomers or dimers during the ruptures we used the 'burning' algorithm introduced by Herrmann et al.\  \cite{HerrmannPRL84} to study the internal structure of percolating  clusters on square lattices. This algorithm detects the presence of single connected bonds (also called 'red' bonds) in percolating clusters and networks. Monomers and dimers are composed by atoms that are equivalent to 'red' bonds if we consider the nanowire as a network of connected atoms. Classically, if we eliminated these particular atoms from the nanowire there will be no conduction between the external nanowire supporting slabs. We defined that two atoms are connected when the distance between their centers is lower that a quantity that we have called the conduction distance $d_C$. In particular, in all results reported here we used $d_C = 3 \AA$. We tested conduction distances closer to the nickel nearest neighbors distance but not relevant differences between results were observed. In our study we have applied the 'burning' algorithm every 1000 MD steps in order to save computational time and, simultaneously, define a reasonable statistical ensemble.

\section{Results and Discussion}

\subsection{Minimum cross-section histograms}

\begin{figure}[tb]
\includegraphics{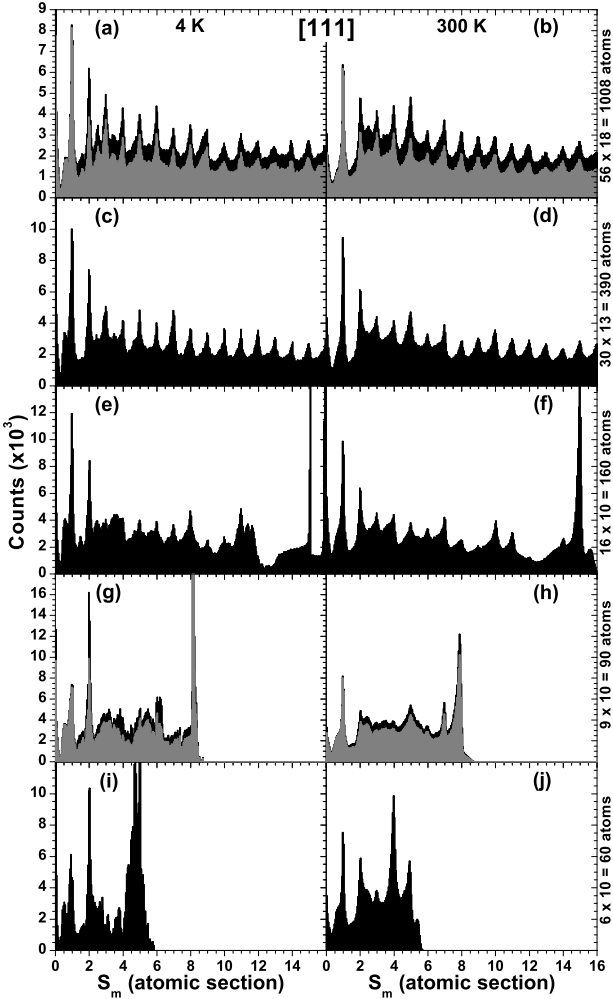}
\caption {\label{fig3} Minimum cross-section histograms $H(S_{m})$ (black filled curves) built with 300 independent Ni nanowire ruptures under stretching along the [111] direction at T=4 K (a,c,e,g,i) and 300K (b,d,f,h,j). Different rows correspond to different initial parallepides sizes as indicated by the right side labels (atoms per layer $\times$ number of layers). 
For two nanowire sizes (1008 and 90 atoms) we also show the partial decomposition of the minimum cross-section histograms $H(S_m)$. Gray filled histograms were constructed taking into account those samples that showed monomers or dimers in the interval $0.25 < S_m < 1.75$. 
}
\end{figure}

\begin{figure}[tb]
\includegraphics{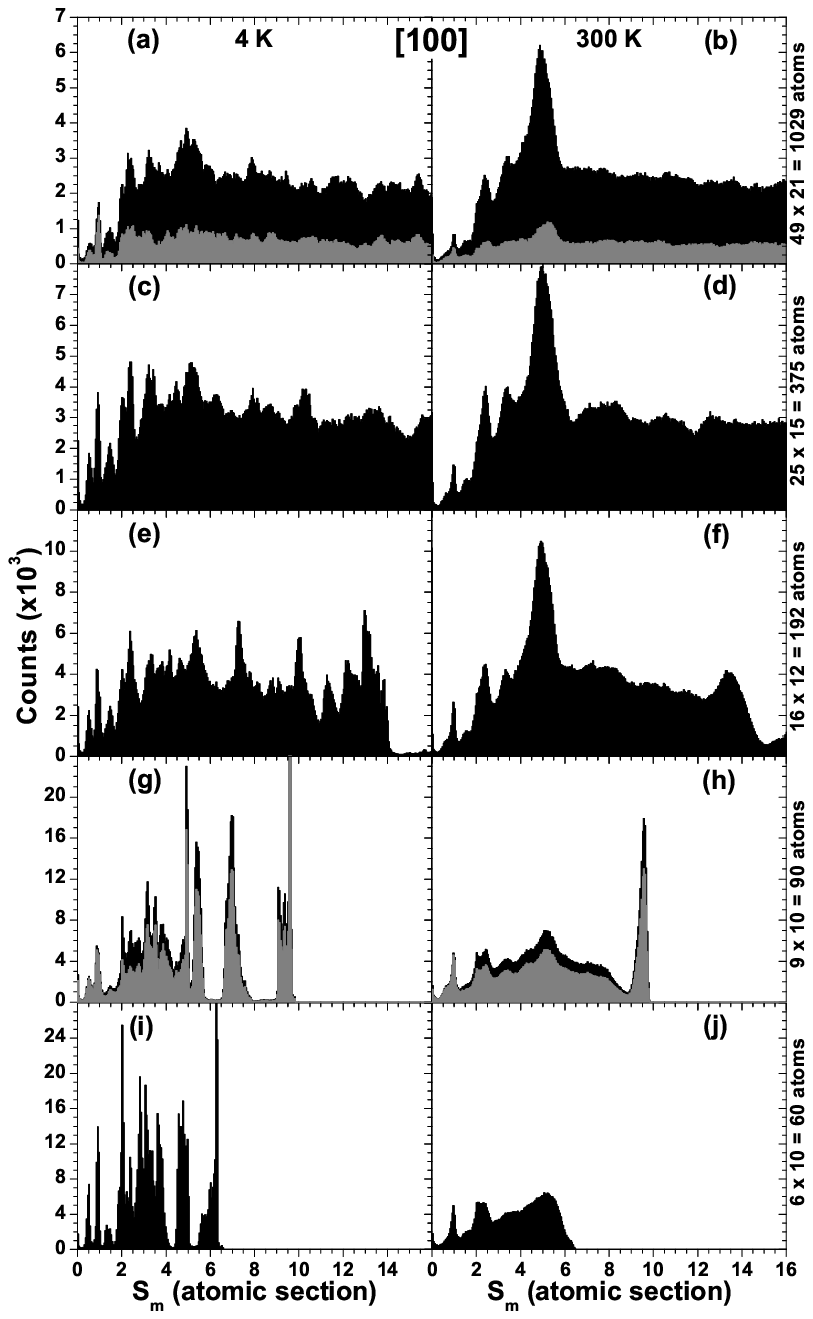}
\caption {\label{fig4} Minimum cross-section histograms $H(S_{m})$ (black filled curves) built with 300 independent Ni nanowire ruptures under stretching along the [100] direction at T=4 K (a,c,e,g,i) and 300K (b,d,f,h,j). Different rows correspond to different initial parallepides sizes as indicated by the right side labels (atoms per layer $\times$ number of layers). 
For two nanowire sizes (1029 and 90 atoms) we also show the partial decomposition of the minimum cross-section histograms $H(S_m)$. Gray filled histograms were constructed as described in Fig.\ \ref{fig3}.
}
\end{figure}

\begin{figure}[tb]
\includegraphics{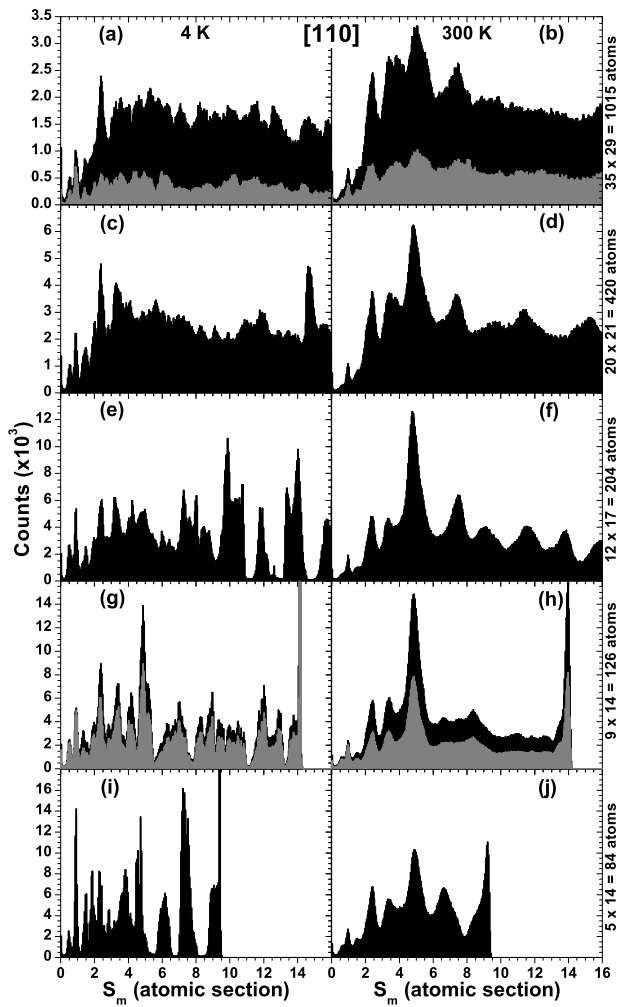}
\caption {\label{fig5} Minimum cross-section histograms $H(S_{m})$ (black filled curves) built with 300 independent Ni nanowire ruptures under stretching along the [110] direction at T=4 K (a,c,e,g,i) and 300K (b,d,f,h,j). Different rows correspond to different initial parallepides sizes as indicated by the right side labels (atoms per layer $\times$ number of layers). 
For two nanowire sizes (1015 and 126 atoms) we also show the partial decomposition of the minimum cross-section histograms $H(S_m)$. Gray filled histograms were constructed as described in Fig.\ \ref{fig3}.
 }
\end{figure}

Minimum cross section histograms $H(S_{m})$ have been built by accumulating $S_{m} (t)$ traces acquired during the simulation of hundreds of nanowire stretching and breaking processes. In Figs. \ref{fig3}, \ref{fig4} and \ref{fig5} we show the histograms $H(S_{m})$, at 4 K and 300 K, for the [111], [100] and [110] stretching  directions, respectively. Each histogram has been constructed with 300 independent breakages. The different histograms are depicted in the range $0 < S_m < 16$. Note that this range includes the whole evolution of those $S_m (t)$ traces corresponding to the small size nanowires. These histograms clearly show the effect of temperature, initial nanowire size, and crystallographic stretching directions on the average structural evolution of breaking Ni nanowires.

A first inspection of these figures reveals the existence of well defined peaks associated to preferred nanowire configurations as it has been shown in previous works\cite{HasmyPRL01,DiazNanotech01,MedinaPRL03,HasmyPRB05,GarciaMochalesAPA05}. A first general feature we can easily identify is the effect of temperature: peaked structures are rather sharp at T=4K whereas they present more rounded shapes at T=300K. Small peaks at T=4K correspond to metastable configurations with slightly higher cohesive energies with respect to other metastable configurations. The increase of temperature allows the exploration of more configurations during the stretching process, and, in this way, those metastable configurations with local minimum energy are easily accessible, leading to a better definition of their associated $H(S_m)$ peaks. 

We have observed, in general, that the temperature increases preserves the general shape of Ni $H(S_m)$ histograms leading to a better definition of the peaked structure. This allows, for instance, an easier identification of those 'magic' configurations\cite{HasmyPRL01,MedinaPRL03,GarciaMochalesAPA05}. The thermal effects are more remarkable for [100] and [110] stretching directions where rounding effects of the peaked structure, and the disappearance of small peaks, become more evident. The more noticieble difference between low and room temperature histograms correspond to the [100] and [110] cases. For both orientations, histograms present an protruding $S_m=5$ peak at T=300K. We will discuss in the next subsection the structure of this peak which is not developed at low temperatures.

A second streaking feature is that the overall histogram shape is modified when the initial nanowire cross-section decreases. The first change is the obvious disappearance of high $S_m$ peaks as the initial nanowire cross-section takes smaller values. This happens because large $S_m$ values are not accessible for small initial section nanowires. The histogram $H(S_m)$ also changes in the low $S_m$ region for decreasing initial cross-sections $N_s$ (see Table \ref{table1}). This change is more remarkable for the [111] histograms (see Fig.\ \ref{fig3}) which are characterized by $H(S_m=1) > H(S_m=2) > H(S_m =3)$ at T=4K for initial cross sections $N_s < 16$. However this decreasing peaked structure changes at T=4K when $N_s < 9$. At higher temperatures (T=300K) the low cross-section region of the [111] histogram slightly changes when the initial nanowire thickness decreases, keeping the trend $H(S_m=1) > H(S_m=2) > H(S_m =3)$. A similar situation happens for [100] and [110] histograms: effects due to the initial nanowire sections are more evident at low temperatures. At room temperature $H(S_m)$ histograms present similar peaked structures  in the small $S_m$ region ($S_m < 2$).

\begin{table}[tb]
\begin{ruledtabular}
\begin{tabular}{|ccc|ccc|ccc|}
 & [111]& &  & [100] & & & [110] &  \\
\hline
$N_{a}$	& 4K		& 300K 	& $N_{a}$ 	& 4K 	& 300K 	& $N_{a}$ & 4K 	& 300K  \\ 
\hline
1008 	& 70697 	& 56479 	& 1029 		& 21230	& 13559	& 1015 	& 16881 	& 12263 \\
 390		& 92298	& 77267 	&  375 		& 47088 	& 20827 	&  420 	& 31257 	& 16174 \\ 
 160		& 106429 	& 92363 	&  192 		& 52253 	&  30228 	&  204 	& 57087 	& 23653 \\
  90   	& 94387  	& 74977  	&   90   		& 56393  	& 52034  	&  126   	& 65007  	& 33111 \\
  60   	& 68770  	& 75046 	&   60   		& 82846  	& 53489 	&   84  	& 90280 	& 40277 \\
\end{tabular}
\end{ruledtabular}
\caption {\label{table2} Number of configurations ($N_{conf}$) with minimum cross section $S_m$ in the interval $0.25 < S_m < 1.75$, found in the corresponding set of 300 independent nanowire breaking events. $N_{conf}$ is shown for [111], [100] and [110] stretching directions, five nanowire sizes $N_{a}$ (see Table \ref{table1}) and two temperatures.}
\end{table}

Therefore, we have found that $H(S_m)$ histograms are very dependent on the stretching (i.e the nanowire axis) direction. Whereas the [111] direction provides histograms with well defined decreasing peaked structure in the low $S_m$ region, the situation dramatically changes for [100] and [110] directions. These two orientations presents $H(S_m)$ histograms with a clear 'depletion' in the region $S_m < 2$. This is more evident when we analyze the total number of breaking events ($N_{conf}$) we have recorded in the interval $0.25 < S_m < 1.75$. This quantity corresponds to the area below the histograms curve in this $S_m$ range. The quantity $N_{conf}$ is shown on Table \ref{table2} for all the stretching directions, temperatures and initial nanowire geometry. 

It is worth see that the temperature determines two different behaviors of $N_{conf}$ as a function of the nanowire size. At low temperatures (T=4K), for large initial nanowire cross-sections, $N_{conf}$ is larger for the  [111] nanowire directions when compared to the values found for the other two studied directions, [100] and [110]. However, for the smallest initial nanowire cross-sections, the  situation reverses and $N_{conf}$ is larger for [100] and [110] directions. At room temperatures (T=300K), the quantity $N_{conf}$  obtained for [111] nanowire directions takes larger values than those obtained for [100] and [110] nanowires in the whole range of initial nanowire sizes we have considered. 

The dependence of $H(S_m)$ on the nanowire size and orientation is a very important issue since theoretical conductance histograms $H(G)$  will also reflect this dependence. In our particular case (nickel), the use of the [100] stretching direction as well as narrow initial nanowires will determine a conductance histogram with small contributions from the low $S_m$ region (narrower nanocontacts), and in particular with low contribution of those peaks located within the region $0.25 < S_m < 1.75$ (mainly formed by monomers and dimers as we will show later on this study). On the contrary, the [111] stretching direction and a thick initial nanowire provide a different conductance regime with large contributions coming from the region $0.25 < S_m < 1.75$. The important point we would like to remark is that these differences explain the apparently conflicting histograms $H(S_m)$ found in the literature. It is clear that calculated histograms corresponding to the [111] stretching direction, at low and room temperatures, and large initial sections\cite{GarciaMochalesAPA05} (see Fig.\ \ref{fig3}(a,b)) are rather different to that calculated for the [100] stretching direction, at T=4K, and small initial sections\cite{PaulyPRB06} (see Fig.\ \ref{fig4}(i)).
 
Finally,  we believe that, whatever it is the stretching direction, computational histograms obtained from large initial section nanowires must be taken into account in order to explain experimental results, where usually initial cross sections (i.e. conductance values) are very large. On the contrary, those histograms obtained from small initial cross section simulations should be carefully handled since they do not capture the behavior of actual thick nanonecks.

\subsection{Analyzing the peak at $S_m \sim 5$ for [100] and [110] stretching directions: pentagonal Ni nanowires}

\begin{figure}[tb]
\includegraphics{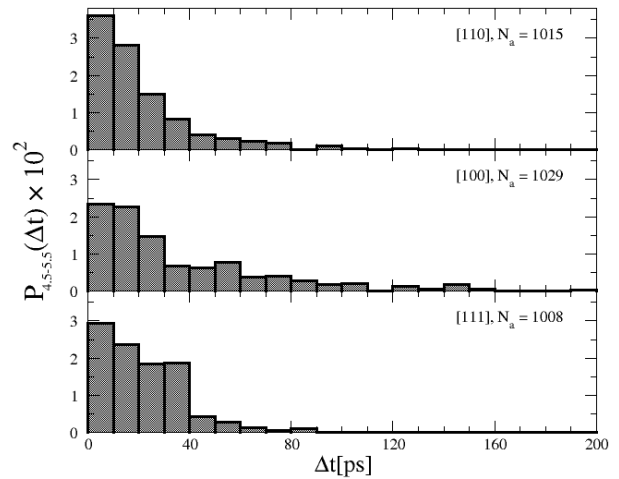}
\caption {\label{fig6} Distribution function $P_{4.5-5.5}(\Delta t)$ at $T=300\mbox{K}$ of the time spent by the largest diameter nanowires during their breaking process within the minimum cross-section interval $4.5 < S_m <5.5$, for the three stretching directions: [111] (bottom), [100] (middle), and [110] (top).  }
\end{figure}

We have pointed out in the previous subsection that minimum cross-section histograms $H(S_m)$ present a huge peak around $S_m = 5$ for the [100] and [110] cases at $T=300\mbox{K}$ (see Figs. \ref{fig4} and \ref{fig5}). If we restrict our study to thicker nanowires, it is clear that the largest well defined peak at $S_m=5$  corresponds to the [100] case. We have carefully analyzed the different nanowire configurations which contribute to the formation of this peak. First, we have calculated the distribution function $P_{4.5-5.5}(\Delta t)$ of the time $\Delta t$ lasted by the nanowire between $S_m=4.5$ and $S_m=5.5$ for the three stretching directions at $T=300\mbox{K}$. Results corresponding to $P_{4.5-5.5}(\Delta t)$ are shown in Fig.\ \ref{fig6}. It is clear that the distribution is wider for the [100] case, in correspondence with the information provided by $H(S_m)$ histograms. Notice that we have found  some $S_m (t)$ traces where the nanowires stays at the $4.5 <  S_m < 5.5 $ region for $\Delta t \sim 200$ ps.

\begin{figure}[tb]
\includegraphics{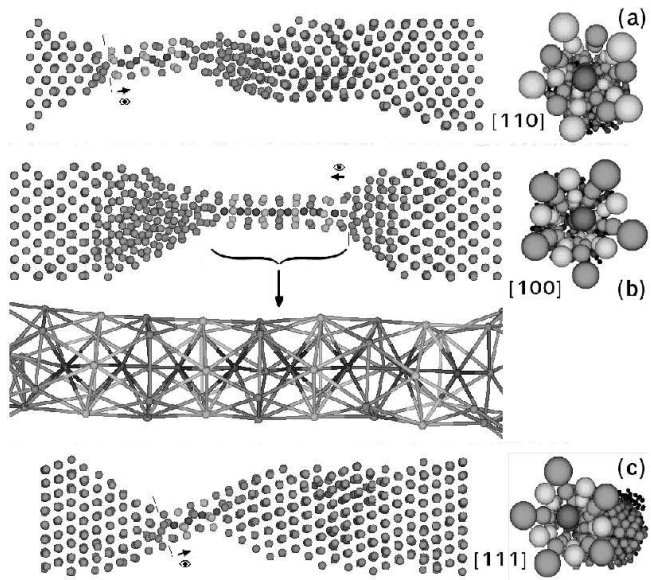}
\caption {\label{fig7} Snapshots of three configurations illustrating the appearance of long wire structures with atomic section close to $S_m \sim 5$ for the [110] (a), [100] (b), and [111] (c) stretching directions. These configurations correspond to the largest initial nanowire cross-sections we have considered (see Table \ref{table1}). The rigth column shows cross-sectional perspective views as seen from the position indicated on the right column figures. The atomic configuration at the central part of the nanowire shows the staggered pentagonal structure -5-1-5-1- discussed on the text.}
\end{figure}

Our MD simulations allow to monitor the evolution of the full set of atomic coordinates during the breaking process. In particular, we have depicted in Fig.\ \ref{fig7} some examples of those nanowires with a relatively long staying-power time in the $4.5 < S_m < 5.5$ region, defining the pronounced peak at $S_m =5$, for the [100], [110] and [111] cases. Note that the origin of the huge peak is related to the tendency of the system to form a long wire with $S_m \sim 5$. At this point we should mention that these long structures are very common for [100] and [110] stretching directions, whereas rarely occur for the [111] case. In general, we have found that those nanowires forming long structures break without a progressive diminishing of its atomic section, i.e. the system does not form monomers or dimers just before the rupture. On the contrary, when the nanowire forms a long structure, it usually breaks following the rupture pattern we have described as "others" (see Fig.\ \ref{fig2}). 

A closer inspection to Fig.\ \ref{fig7} allows to determine that these depicted wires present a well defined sequence of the atoms $\cdots -5-1-5-1-\cdots$. This sequence does not correspond to any crystallographic fcc or bcc structure. This type of arrangement is not seen at 4K because a larger temperature is required to explore and overcome those energy barriers leading to  configurations able to develop these pentagonal chains. We have found that these pentagonal nanowires are formed by subsequent staggered parallel pentagonal rings (with a relative rotation of $\pi /5$) connected with single atoms.  A similar, but shorter, structure was found for Cu(111) breaking nanowires using MD simulations\cite{MehrezPRB97}.  The stability of such pentagonal Cu nanowires was later confirmed by ab-initio calculations\cite{SenPRB02}, demonstrating that staggered pentagonal nanowires are favorable configurations. More recently, MD tight-binding calculations have determined the presence of such pentagonal -5-1-5-1- nanowires during the breaking process\cite{GonzalezPRL04} of [110] Cu nanowires.  Pentagonal motives also appear in infinite Al nanowires\cite{GulserenPRL98}.  

The most relevant point is that we have statistically demonstrated that the RT stretching of Ni nanowires along the [100] or [110] directions provide a method to generate staggered pentagonal nanowires with relative high probability.

\subsection{Analysis of the formation of monomer and dimer configurations}


The study of the minimum cross-section interval $0.25 < S_m < 1.75$ is important because it defines the set of atomic configurations that present the type of structures depicted in Fig.\ \ref{fig2}, i.e. monomers, dimers and more complex structures. The electron transport during the last nanowire breaking stages strongly depends on the type of nanowire atomic arrangements characterizing these final stages. 
Therefore it is critical to know the type of involved atomic configurations (monomers, dimers, other) appearing in the interval $0.25 < S_m < 1.75$ (around the $S_m = 1$ histogram peak) as well as their statistical weights. We have analyzed these statistical weights for three stretching directions ([111], [100] and [110]), two temperatures,  and several nanowire sizes.

\begin{table}[tb]
\begin{ruledtabular}
\begin{tabular}{|c|c|c|c|c|}
 stretching direction & $N_{a}$ & T (K) & $N_{mon}$ & $N_{dim}$ \\
\hline
[111]	&1008	&4		&344		&187 \\ 

 		&		&300		&270		&163 \\

\hline
[100]	&1029	&4		&76		&48 \\ 

		&		&300		&46		&40 \\
\hline
[110]	&1015	&4		&51		&37 \\

		&		&300		&60		&29 \\
\end{tabular}
\end{ruledtabular}
\caption {\label{table3} Total number of monomers ($N_{mon}$) and dimers ($N_{dim}$) analysed with the Herrmann\cite{HerrmannPRL84} algorithm for the largest studied nanowires (with size $N_{a}$), and for [111], [100] and [110] stretching directions, at T=4K and T=300K.}
\end{table}

The geometric characterization of the $0.25 < S_m < 1.75$ region  has been accomplished using the Herrmann 'burning' algorithm every 1000 MD steps. Note that the number of configurations we analyze is smaller than the number of data used to construct the $H(S_m)$ histogram. As an example, Table \ref{table3} shows the number of configurations where dimer or monomer have been detected for the three largest nanowires corresponding to [111], [100] and [110] stretching directions, respectively. A first inspection of Table \ref{table3} reveals that the number of situations with monomer and dimers is approximately five times larger for the [111] case than for the [100] and [110] cases. This means that, for the [111] case, the nanowire breaks gradually from more complex structures to monomers which, in some situations, form dimers. For the other two stretching directions there are many configurations that directly break from situations with minimum cross-section $S_m > 2$.  In consequence, the $H(S_m)$ peak at $S_m=1$ is the largest one for the [111] case, whereas this peak is considerably smaller than those peaks formed at $S_m > 2$ for the other two stretching directions.  Interestingly, we have found for smaller initial system sizes that the number of configurations with monomers and dimers is similar for the three stretching directions.


\begin{figure}[tb]
\includegraphics{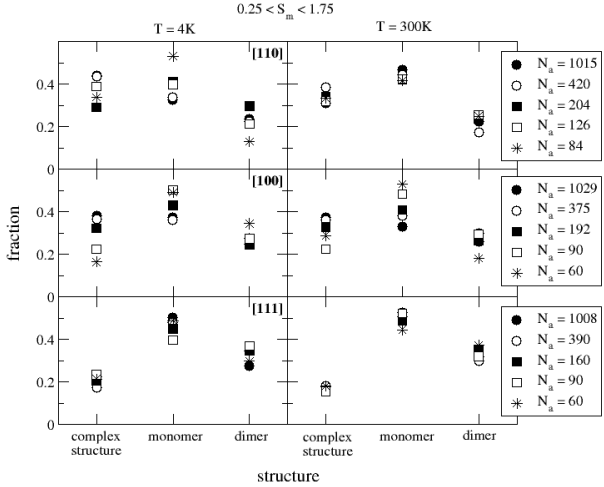}
\caption {\label{fig8} Fraction of monomer, dimers and more complex structures appearing during the Ni nanowire breaking process within the range $0.25 < S_m < 1.75$. }
\end{figure}

The fractions of monomers, dimers and complex structures appearing in the range $0.25 < S_m < 1.75$ are depicted in Fig.\ \ref{fig8}. For the [111] stretching direction we note that the join proportion of monomer and dimers ($\sim 80$\%) is larger than the fraction of other structures ($\sim 20$\%). This fraction seems to be rather independent on the temperature as well as the initial nanowire size. This is a very interesting result since it implies that final configurations of a breaking Ni nanowire under [111] stretching mainly correspond to monomer and dimer structures. 

For the other two stretching directions [100] and [110], the fraction of other (more complex) structures increases for both temperatures and also shows a larger dependence on the nanowire size. In some cases, the fraction of complex structures takes values close to 40\%.  In addition we have found that the fraction of monomers is rather similar to that found for the $[111]$ case. However, this is not the case for the fraction of dimers, which takes, in some specific cases, values below 20\%. From Fig.\ \ref{fig8} we also infer that there is less data dispersion for the [111] case than for the other two stretching directions. Note that part of this dispersion may be caused from the use of a poor statistical description for the [100] and [110] cases.


\begin{figure}[tb]
\includegraphics{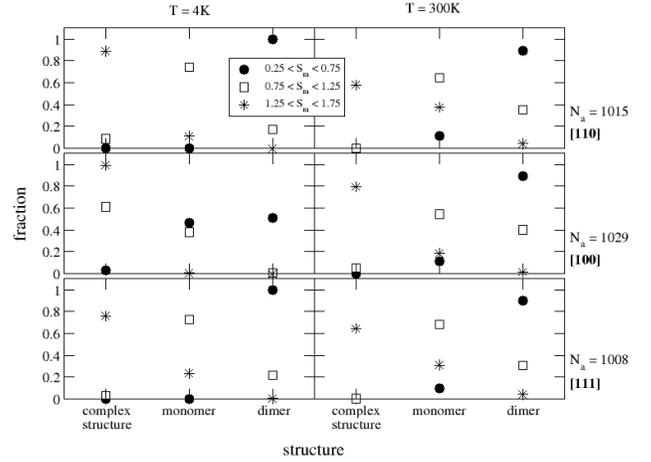}
\caption {\label{fig9} Fraction of monomers, dimers and more complex structures that are observed around the sub-peaks located at $S_m=0.5$, $1.0$ and $1.5$ in the $H(S_m)$ histograms. Results for $T=4K$ (left) and $T=300K$ (right) are shown for the largest systems simulated for each one of the stretching directions (see Figures \ref{fig3}(a,b), \ref{fig4}(a,b) and \ref{fig5}(a,b)). }
\end{figure}

When we analyze the minimum cross-section range $0.25 < S_m < 1.75$ it is worth see that minimum cross-section histograms $H(S_m)$ do not present a simple structure with a single well-resolved peak centered at $S_m =1$. On the contrary, in many cases (see Figs. \ref{fig3} - \ref{fig5}) we identify the presence of satellite peaks around $S_m = 0.5$ and $S_m = 1.5$. We have explored such peaks to determine the type of atomic configurations included on them. In Fig.\ \ref{fig9} we show the probability of finding different structures (monomer, dimer, and complex structures) around each one of the sub-peaks we have commented above. For the sake of simplicity we have considered, for T=4K and T=300K, the largest nanowire sizes defined for the three stretching directions we have considered. In general, we notice that the peak at $S_m = 0.5$ is basically originated from dimers whereas the peak at $S_m = 1$ comes from a mix of monomer and dimer configurations. Finally,  the region around $S_m = 1.5$ is formed with contributions from complex structures and a small proportion of monomers.  This general trend is not observed at $T=4\mbox{K}$ for the [100] stretching direction. In this case, the difference could arise from statistical errors coming from the short time that the system spends in the $S_m \sim 1$ region.

In the analysis of Figs. \ref{fig8} and \ref{fig9} we have argued that the small number of detected configurations with the Herrmann "burning" algorithm is a key ingredient to understand the dispersion of values found when analyzing the probabilities of finding monomer, dimer or more complex configurations. This number depends on both the detection window (1000 MD steps) and the intrinsic breaking times which could, in principle, depend on the stretching direction, temperatures and the nanowire initial size. In order to get insight about the dynamics of the last stages of the nanowire we analyze the distribution of times associated to the breaking process. To do that we have analyzed the distribution function $P_{0.25-1.75}(\Delta t)$ of the time $\Delta t$ spent by the nanowire in the interval $0.25 < S_m < 1.75$. This distribution is similar to the plateau lengths distribution used to experimentally detect linear atomic chains\cite{YansonNature98}.

\begin{figure}[tb]
\includegraphics{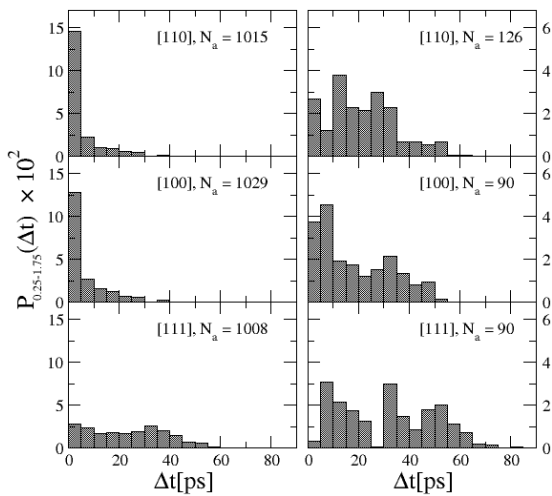}
\caption {\label{fig10} Distribution function $P_{0.25-1.75}(\Delta t)$ of the time spent by the set of largest (left) and mid-size (right) Ni nanowires (at $T=4\mbox{K}$) within the interval $0.25 < S_m <1.75$, for the stretching directions [111] (bottom), [100] (middle) and [110] (top). }
\end{figure}

In Fig.\ \ref{fig10} we show $P_{0.25-1.75}(\Delta t)$ for the largest nanowires we used for each one of the stretching directions as well as for three mid-size nanowires. Note that for the [100] and [110] cases a huge peak is observed below 5 ps. The distribution falls down fast as the time interval $\Delta t$ is increased. That is, the system spends a short time in this minimum cross section region. For the [111] case we found a different dynamical pattern: the distribution broadens showing that the nanowire forms stable monomer and dimer-like configurations
and giving rise to the strong peaked $H(S_m)$ structure depicted in Fig.\ \ref{fig3}.  
When the initial system size decreases the behavior for the three stretching directions is rather similar. In particular, the presence of two peaked structures in $P_{0.25-1.75}(\Delta t)$ could correspond directly to the formation of monomer (one atom chain) and dimers (two atoms chain), as it has been found for some metallic species\cite{YansonNature98}. Therefore, it is important to remark that the system dynamics during the last breaking stages strongly depends on the nanowire stretching direction and the system size.

We have also addressed the question of the origin of the atoms forming the monomer and dimer configurations, in a similar way to previous studies done for gold nanowires\cite{SatoAPA05}. We have performed the statistical analysis for all the nanowires we have considered in order to determine whether those atoms involved in the formation of the narrowest nanocontact region, at the last breaking stages, come from bulk or surface positions. For every nanowire we have followed back the atomic trajectories of those atoms found in monomer and dimer configurations in order to find its initial position after the stretching process. We have found that surface (bulk) atoms contribute to the formation of monomers with a statistical weight very similar to the initial fraction of surface (bulk) atoms. Therefore there is not any relevant role for the surface atoms in the formation of monomers and dimers, at variance of results found for gold\cite{SatoAPA05}. The origin of this discrepancy may arise from the comparison between two different metallic species (remember that, for instance, gold forms long atomics chains whereas nickel does not) or from the type and quality if the statistical analysis (different number of studied breaking events).

\subsection{Histogram decomposition into monomer and dimer components}

The decomposition of conductance experimental histograms into partial histograms has been already proposed\cite{RegoPRB03} to study the influence of the nanowire crystalline orientation on the conductance traces or to determine the presence of magnetic and non-magnetic contributions on Ni and Cu conductance histograms\cite{DiazJMMM06}. From a theoretical perspective, the decomposition on partial histograms is an additional tool that allows to understand the dynamical evolution of the set of nanowires taken into consideration to build up the full histogram\cite{DreherPRB05,HasmyPRB05,PaulyPRB06}. 

In Figs.\ \ref{fig3}, \ref{fig4}, and \ref{fig5} we plotted the minimum cross-section histograms $H(S_m)$ for both temperatures and several nanowire sizes. We have decomposed this histogram for two nanowire  sizes. In black we show the histograms using the complete set of breaking events. In gray we show the partial $H(S_m)$ histograms constructed from those configurations that present monomer and dimer configurations in the interval $0.25 < S_m < 1.75$. 

For the [111] stretching direction cases (see Fig. \ref{fig3}(a,b,g,h)), we notice that the decomposed histograms are quite similar -in all the $S_m$ range- to the full histogram for both temperatures and both system sizes. This indicates that, for the [111] stretching direction, all the traces $S_m (t)$ evolve from large $S_m$ values until breaking using monomer and/or dimer breaking patterns. This breaking scheme is very robust and seems to be independent on the initial nanowire size and temperature (for temperatures below RT).

However, for the other two stretching direction cases ([100] and [110]) there are strong differences between large and small nanowire sizes. For the largest studied systems (see Figs. \ref{fig4}(a,b) and \ref{fig5}(a,b)), we notice that the monomer and dimer structures are solely responsible of the formation of those smaller peaks located at the region $0.25 < S_m < 1.75$ but there is a lot of configurations with large $S_m$ values that do not break forming monomers neither dimers.  In this case (large system sizes), we also see that many $S_m(t)$ curves forming long pentagonal nanowires (giving rise to the huge peak observed around $S_m \sim 5$) do not break using the monomer or dimer breaking pattern and directly break from relatively large cross sections. The temperature does not play any relevant role in the studied temperatures range (except the rounding and smoothing of the histogram peaks). 

For smaller nanowires that break under [100] and [110] stretching (see Figs. \ref{fig4}(g,h) and \ref{fig5}(g,h)) we recover the breaking pattern we obtained for the [111] small size case, where almost all $S_m(t)$ traces present well defined monomer or dimer configurations during the last stages of the elongation process. 

Therefore, the relative formation of monomers and dimers during breaking events strongly depends on the system size when we elongate the nanowire in the [100] and [110] directions. The [111] stretching direction do not show important changes in the formation of monomer-dimer configurations for decreasing nanowire sizes. If we assume that the stretching and breaking of large size simulated nanowires are closer to the experimental situation, we clearly see that the appearance of monomers and dimers (and the shape of $H(S_m)$) will strongly depend on the experimental stretching direction used to elongate and break the nanowire or nanocontact.

\subsection{Local environment of monomer and dimer breaking configurations}

It is obvious that the local environment is a key factor to determine the conductance of monomer or dimer configurations. This has been demonstrated for Al nanocontacts where monomer and dimer transport properties present different features\cite{DreherPRB05,HasmyPRB05}. For Al dimer-like configurations, the interaction between supporting atomic structures surrounding the dimer becomes negligible and the conductance is governed by three electron channels whose transmission coefficients determine a conductance value close to $G_0$.\cite{DreherPRB05,ScheerPRL97} On the contrary, for a monomer configuration the distance between atomic supports decreases, enhancing their orbital interaction, and finally leading to an incrase of the conductance upto  $G \sim 2G_0$. We can guess a similar situation for Ni, where $d$ orbitals associated to supporting pyramid-like structures will interact strongly for the monomer situation giving rise to higher conductance values whereas the interaction will decrease for the dimer-like configuration, leading to lower conductance values. Static conductance calculations on Ni configurations (monomer and dimer)\cite{SirventPRB96} are consistent with that picture.

\begin{figure}[tb]
\includegraphics{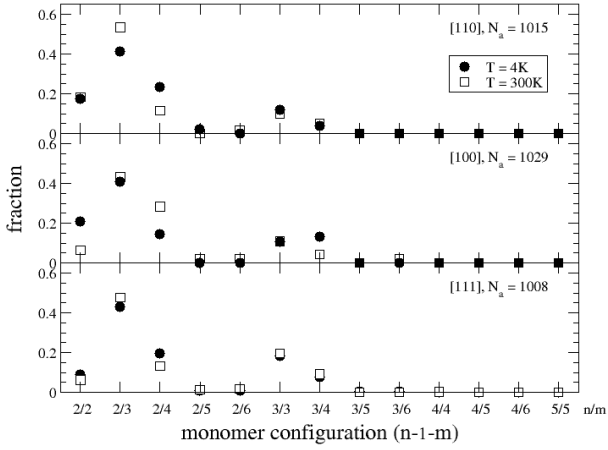}
\caption{\label{fig11} Probability of occurrence of different atomic environments around a Ni monomer. A $n/m$ configuration  means that the monomer configuration is of type "n-1-m" or "m-1-n", i.e. there are $n$ atoms at one side of the monomer and $m$ atoms at the other side. The probability has been calculated for the largest nanowire used for the stretching directions [111] (bottom), [100] (middle) and [110] (top). Two temperatures have been studied: 4K (black dots) and 300K (squares).}
\end{figure}

\begin{figure}[tb]
\includegraphics{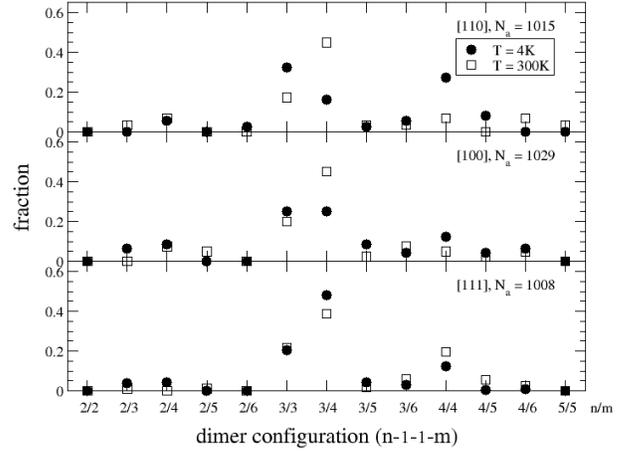}
\caption{\label{fig12} Probability of occurrence of different atomic environments around a Ni dimer. A $n/m$ configuration  means that the dimer configuration is of type "n-1-1-m" or "m-1-1-n", i.e. there are $n$ atoms at one side of the dimer and $m$ atoms at the other one. The probability has been calculated for the largest nanowire used for the stretching directions [111] (bottom), [100] (middle) and [110] (top). Two temperatures have been studied: 4K (black dots) and 300K (squares).}
\end{figure}

From the previous analysis it is obvious that the local atomic environment will play an important role in the conductance determination, specially in the monomer case. MD calculations allow to establish the type of neighborhood formed around monomer and dimer configurations. In Figs.\ \ref{fig11} and \ref{fig12} we present the occurrence probability of different atomic configurations around monomers and dimers. Histograms for different temperatures and stretching directions have very similar shape. We can conclude that the atomic structures (and its probability of occurrence) around monomers and dimers are roughly the same independently on the stretching direction and the temperature. In Fig. \ref{fig11} we see that the most probable monomer structure presents the configuration 2-1-3. Around 40\% of the monomers exhibit this configuration. But this is not the only important configuration since we found that 2-1-4, 3-1-3, 3-1-4 and 2-1-2 configurations for all the stretching directions appear with a probability between $\sim 20$\% and $\sim 10$\%. Fig.\ \ref{fig12} presents the analysis of the local environment for those dimers we have detected in our MD simulations. It can be observed that the most common dimer configurations are 3-1-1-4, 3-1-1-3, and 4-1-1-4 with ocurrence probability between $\sim 40$\% and $\sim 20$\%. In addition, some other configurations can be observed with probability below 10\%. 

It is clear that in order to analyze the experimental conductance histograms it is important to take account the conduction coming for all these atomic configurations,  including the adequate proportion of monomer and dimers as well as the correct proportion of possible atomic environments for each monomer and dimer situation. The present information  is very practical to address, for instance, the ab-initio study of the electronic transport through static monomer or dimer-like Ni structures. In general, the non-trivial 2-1-3 configuration would have little opportunities to be included as the most likely monomer structure, whereas past calculations only studied 3-1-3\cite{SirventPRB96} and 4-1-4\cite{SirventPRB96,JacobPRB05} configurations. In particular, we have not found evidences of the structure 4-1-4 in our simulations. 

\subsection{Non-local environment of monomer and dimer breaking configurations}

The accurate knowledge of the local environment of the monomer and dimer configurations is not enough to understand the structure of conductance histograms. Electron transport mainly depends on the narrowest nanocontact cross-section and the surrounding regions. However, the electron transport is not only governed by the small number of open channels allowed by the nanoneck geometry. In fact, the transmission coefficients appearing  in Eq. (\ref{landauer1}) depend on several factors including the possible presence of disorder (defects, vacancies, impurities, dislocations, ...) along the wire. 

\begin{figure}[tb]
\includegraphics{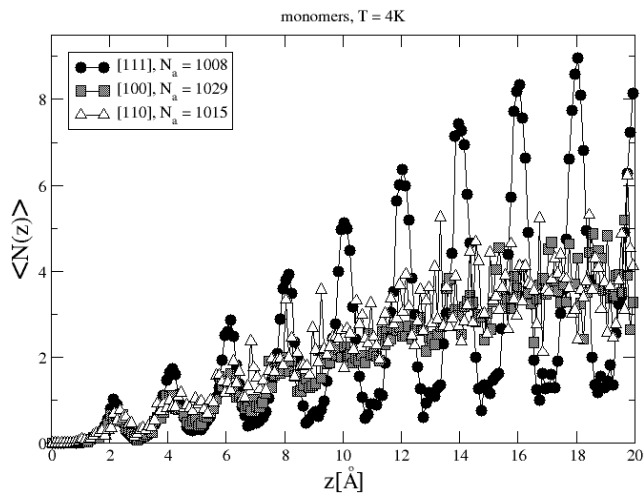}
\caption{\label{fig13} Average number of atoms $<N(z)>$ found at a distance $z$ from the monomer position, at $T=4$K, for the stretching directions [111] (bottom), [100] (middle), and [110] (top).}
\end{figure}

A first approach to unveil the existence of disorder along the nanowire is to determine the atomic structure as the $z$ distance from the monomer (or dimer) position increases.  In Fig.\ \ref{fig13} we plot the average number of atoms $<N(z)>$ located at a distance $z$ from the monomer position for the largest systems we have simulated. We have calculated the average number of atoms for the three stretching directions, at $T=4K$, and taking into account all the monomer configurations appearing during 300 breaking events. On one hand, for the [111] case, we note clear peaks separated a distance close to the typical separation of the [111] planes $d_{111} = 2.03 \AA$. This points that the system keeps the same crystallographic structure from the fixed slabs, at both sides of the sample, towards the monomer position. On the other hand, for the [100] and [110] stretching directions, the situation is quite different. Both orientations present a very similar local atomic structure for monomer and dimers. This local structure also agrees with that observed for the [111] case (as described in the previous subsection). However when $z$ increases we could not distinguish a peaked structure similat to that of the [111] case. For these cases, the fixed slabs try to maintain the [100] or [110] structure but the nanowire presents dislocations, disorder, and the formation of  large non-crystalline wires when we study regions closer to the monomer.  A very similar behavior is observed around dimmers (not shown).

This kind of situation is not exclusive of monomer and dimer configurations appearing in [100] and [110] cases. In fact, this lack of ordering happens during the whole stretching process. An appealing example appears in Fig.\ \ref{fig7}. The unusual pentagonal structure found in the [111] stretching direction case (Fig.\ \ref{fig7}(c)) presents a smooth transition towards the [111] supporting frozen bilayers, with little amount of disorder. On the contrary  (Fig.\ \ref{fig7}(a,b)), the pentagonal configurations emerging in the [100] and [110] cases are connected to the frozen bilayers throughout rather disordered structures.

The previously described results are very important whether we try to address the study of the electron transport at the latest stages of the breaking process, i.e. those moments when monomers and dimers govern the number of open channels available for electron transmission. It is clear that the kind of local environment of a monomer (or dimer) does not strongly depend on the stretching direction. Therefore the electron transport should present similar features for all the directions (with different occurrence weights). However, the electron path inside the wire (neglecting inelastic collisions since the typical nanowire size for metals is smaller than the room temperature inelastic mean free path) could  be very different when we compare the [111] stretching direction with the [100] and [110] ones. In the first case ([111]) the electron moves inside a well ordered system, where elastic scattering events are mainly due to the roughness of the nanowire walls. In the second case ([100] and [110] stretching directions) the electron 'feels' the presence of disorder during its motion in a large nanowire region. This disorder will likely originate a number of scattering events leading to a conductance decrease. If the scattering due to disorder becomes very important the electron transport could leave the ballistic regime and quantum diffusive features should appear. Therefore we could observe two opposite transport mechanism depending on the stretching direction. 

These effects possibly explain the computational conductance histograms obtained by Pauly et al. \cite{PaulyPRB06} where a thin nanowire (8 atoms section) was stretched along the [100] direction. Their calculated cross-section and conductance histogram shows that the contribution from monomer and dimer configurations in the region $0 < S_m < 2$ is very small in comparison with the contributions coming from the configurations $S_m > 2$. This is clearly due to the chosen stretching direction (see Fig.\ \ref{fig4}). In addition, their calculated conductances also take values well below $3.5G_0$ for low and relatively large cross-section values (with many open conducting channels due to the presence of $s$ and $d$ orbitals). This decrease of the expected conductance is not only due to the presence of few atoms on the narrowest nanowire part, nor the d-blocking effects, but it points towards diffusive electron transport effects. This behavior is consistent with the presence of a disordered configurations along the nanowire, as expected for [100] stretching (see Fig.\ \ref{fig13}). Therefore we claim that very different conductance histograms will obtained if thick Ni nanowires stretched along the [111] crystallographic direction were used.
 
\section{Conclusions}

We have carried out hundreds of MD simulations of Ni nanowire breaking processes. We use the EAM approach to describe the interatomic many body interaction. From these simulations we are able to follow the evolution on time of the minimum cross-section $S_m (t)$. By adding hundreds of $S_m (t)$ traces we have constructed computational minimum cross-section histograms $H(S_m)$ that statistically unveil the presence of preferred configuration during the elongation and breaking history.  The modification of some initial conditions (temperature, stretching direction or nanowire size) leads to changes on the peaked structure of $H(S_m)$ histograms. In this way, we are able to correlate these histogram changes with changes in the type of favorable structures appearing under stretching.  The last stages of the nanowire breaking process are of special interest since electron transport is determined by a cross section formed by few atoms. However, this situation requires an additional study since the local environment of those atoms forming the nanoneck plays an important role in the electron transport characteristics. Therefore, we have extracted additional information from the $S_m (t)$ traces as the type of configurations that appear at $S_m \sim 1$. We have found that monomers, dimers and other more complex structures are present at the latest stages of the breaking events. We did not find linear atomic chains of three or more atoms for all the systems and stretching directions we analyzed.

We have found that the $H (S_m)$ histograms do not depend dramatically on the temperature within the analyzed temperature range (below 300K). In general, we have only noticed rounding of the peaked structure of $H(S_m)$, the suppression of small $S_m$ fluctuations and the increase, at RT, of the peak located at $S_m =5$ for the [100] and [110] cases. The absence of large thermal effects is consistent with the fact that bulk Ni melting temperature is rather large ($\approx 1728$K), and then RT is not enough to activate additional mechanisms as surface diffusion effects that could modify the nanowire breaking dynamics. 

Summarizing our computational results we have determined three different scenarios for the set of nanowires we have studied. The first scenario corresponds to large nanowires stretched along the [111] direction. In this case we have noticed a well defined series of decreasing $H(S_m)$ peaks as the quantity $S_m$ increases. Nanowires break forming  monomers and dimers, responsible of the large peaked structure found in the $ S_m < 2$ region. We have found that [111] breaking processes present, in general, well ordered structures during the breakage process, and monomer and dimer configurations are surrounded by well defined planes. This non-local ordering with respect the monomer and dimer arrangements preserves the ballistic character of the electron transport.  This scenario is preserved when temperature increases from 4K to RT.

A second scenario corresponds to large nanowires stretched along [100] and [110] directions. In these cases the $H(S_m)$ histogram does not large peaks in the low $S_m$ region ($S_m < 2$). The analysis of such $S_m$ region reveals that monomer configurations also play a relevant in this region. However, the probability of finding dimer-like structures is smaller than that of finding more complex structures. In addition, the long-range environment surrounding monomer or dimer structures is rather disordered, possibly contributing to the conductance decrease. When temperature increases from 4K to RT, we found a dramatic increase of those peaks located at $S_m \sim 5$ for both, [100] and [110] directions. We have confirmed that this peak is caused by the presence of long staggered pentagonal chains with ...-1-5-1-5-... structure.  In relation with the formation of long chains in MD simulations, a recent work\cite{PuJCP07} points towards the higher ductibility predicted by EAM potentials in comparison with other interatomic potentials, leading to the formation of long structures. However, more detailed statistical analysis is required to confirm this behavior.

Finally, we can define a third scenario, common for the three stretching directions, corresponding to small initial nanowire sizes. In this case, $H(S_m)$ histograms become more complicated due to size effects. This noisy scenario is not strongly modified when we increase temperature from 4K to 300K. 

In addition, we have found that monomer and dimer configurations appearing for the three stretching directions present very similar features. In particular we have found that monomers prefer the 2-1-3 and 2-1-4 configurations to the 3-1-3 one. These configurations appear with approximately the same probability for all the stretching directions and temperatures we have considered. Most probable dimer configurations correspond to 3-1-1-4, followed by 3-1-1-3 and 4-1-1-4 arrangements. These results are of importance since they define the kind of static configurations required to describe realistic breaking configurations.

Since we have not determined the actual conductance associated to each atomic configuration it is very difficult to establish a comparison between our computational minimum cross-section histograms and experimental conductance histograms.  Furthermore, the interpretation of experimental Ni conductance histograms is a difficult task since they involve not only mechanical and electrical properties but magnetic ones too. In spite of our model limitations to perform the comparison with experiments we can extract few key points from the present MD simulations. The similar behavior of $H(S_m)$ at 4K and 300K excludes the presence of different atomic configurations as explanation of such differences between low and room temperatures. In addition the presence of well marked $H(S_m)$ peaks is consistent with the appearance of preferred conductance values (giving rise to conductance histogram peaks). We can assume, at low temperatures and clean environment, that experimental histograms present contributions arising from breaking processes corresponding to different crystalline orientations. Thus, it is reasonable to propose that the [111] stretching case provides a strong contribution from monomer and dimer configurations, whereas [100] and [110] do not. In the first case (monomers and dimers) we know that expected conductances\cite{SirventPRB96} fall within the range $G_0$--$2.5G_0$, just in the region where 4K experimental conductance histograms show a broad peak\cite{BakkerPRB02,UntiedtPRB04,CalvoIEEE07}. The second broad conductance peak centered around $G \approx 3G_0$ could be explained in terms of those contributions associated configurations with $S_m < 2$ for the three stretching diretions.

Experimental results found at room temperature cannot be interpreted from the present MD simulation. The complex behavior of Ni conductance histograms noticed at room temperatures must be attributed to different causes: (i) the presence of structural disorder at the nanocontact region\cite{MehrezPRB98}, (ii) the presence of impurities or chemiadsorbed atoms on the nanowire\cite{UntiedtPRB04,LiAPL00}, (iii) ballistic magnetoresistence (BMR) effects originated by abrupt magnetic domain walls anchored at the narrowest cross-section of the nanowire\cite{GarciaPRL99,TataraPRL99,ChopraPRB02,ViretPRB02} or (iv) the increase of the domain wall fluctuations with temperature at the nanocontact region.\cite{SerenaRMF01} Therefore more realistic calculations are required to understand the non convergent experimental results.


\begin{acknowledgments}
We thank J.\ L.\ Costa-Kr\"amer, M.\ D\'{\i}az,  J.\ J.\ Palacios, C.\ Guerrero, E.\ Medina and A.\ Hasmy for helpful discussions. This work has been partially supported by the CSIC-IVIC
researchers exchange program and the Spanish DGICyT (MEC) through Projects FIS2005-05137, BFM2003-01167-FISI, and FIS2006-11170-C02-01, and by the Madrid Regional Government through the Programmes S-0505/MAT/0303 (NanoObjetos-CM) and S-0505/TIC/0191 (Microseres-CM). 
RP also acknowledges Spanish MEC by the financial support through its Researchers Exchange and Mobility Programme. P.G.-M. is supported by the Spanish MEC through its "Ram\'on y Cajal" Program.
\end{acknowledgments}



\begin{thebibliography}{99}

\bibitem{SerenaBook97}  
{\it Nanowires}, edited by P.\ A.\ Serena and N.\ Garc\'{\i}a, 
NATO ASI Series E, Vol.\ {\bf 340} (Kluwer, Dordrecht, 1997);
N.\ Agra\"{\i}t, A.\ Levy-Yeyati, and J.\-M.\ van Ruitenbeek,
Phys. Rep. {\bf 377}, 81 (2003).

\bibitem{LandauerPhilMag70}
R.\ Landauer, Phil. Mag. {\bf 21}, 863 (1970); Z. Phys. B -- Condens. Matter {\bf 68}, 217 (1987); 
J. Phys. Condens. Matter {\bf 1}, 8099 (1989).

\bibitem{PascualPRL93}   
J.\ I.\ Pascual, J.\ M\'endez, J. G\'omez-Herrero, A.M. Bar\'o, 
N. Garc\'{\i}a, and V.\ T.\ Binh,
Phys. Rev. Lett. {\bf 71}, 1852 (1993).

\bibitem{AgraitPRB93} 
N.\ Agra\"{\i}t, J.\ G.\ Rodrigo, and S.\ Vieira
Phys. Rev. B {\bf 47}, 12345 (1993).

\bibitem{OlesenPRL94} 
L.\ Olesen, E.\ Laegsgaard, I.\ Stensgaard, F.\ Besenbacher, J.\ Schi{\o}tz,  
P.\ Stoltze, K.\ W.\ Jacobsen, and J.\ K.\ N{\o}rskov, 
Phys. Rev. Lett. {\bf 72}, 2251 (1994).

\bibitem{OlesenPRL95}  
L.\ Olesen, E.\ Laegsgaard, I.\ Stensgaard, F.\ Besenbacher,
J.\ Scholtz, P.\ Stoltze, K.\ W.\ Jacobsen, and J.\ K.\ Norskov, 
Phys. Rev. Lett. {\bf 74}, 2147 (1995).

\bibitem{MullerPRL92}  
C.\ J.\ Muller, J.\ M.\ van Ruitenbeek and L.\ J.\ de Jongh,
Phys. Rev. Lett. {\bf 69}, 140 (1992);
J.\ M.\ Krans, C.\ J.\ Muller, I.\ K.\ Yanson, Th.\ C.\ M.\ Govaert, R.\ Hesper, and J.\ M.\ van Ruitenbeek, 
Phys. Rev. B {\bf 48}, 14721 (1993).

\bibitem{KransNature95}  
J.\ M.\ Krans, J.\ M.\ van Ruitenbeek, V.\ V.\ Fisun, I.\ K.\ Yanson, and L.\ J.\ de Jongh, 
Nature (London) {\bf 375}, 767 (1995).

\bibitem{KondoPRL97}
Y.\ Kondo and K.\ Takayanagi,
Phys. Rev. Lett. {\bf 79}, 3455 (1997);
T.\ Kizuka, Phys. Rev. Lett. {\bf 81}, 4448 (1998);
H.\ Ohnishi, Y.\ Kondo, and K.\ Takayanagi, 
Nature (London) {\bf 395}, 780 (1998);
V.\ Rodrigues, T.\ Fuhrer, and D.\ Ugarte,
Phys. Rev. Lett. {\bf 85}, 4124 (2000).

\bibitem{KondoSci00}
K.\ Kondo and K.\ Takayanagi, 
Science {\bf 289}, 606 (2000);
Y.\ Oshima, H.\ Koizumi, K.\ Mouri, H.\ Hirayama, K.\ Takayanagi, and Y.\ Kondo,
Phys. Rev. B {\bf 65}, 121401(R) (2002);
Y.\ Oshima, A. Onga, and K.\ Takayanagi, 
Phys. Rev. Lett. {\bf 91}, 205503 (2003).

\bibitem{LiAPL98}  
C.\ Z.\ Li and N.\ J.\ Tao, Appl. Phys. Lett. {\bf 72}, 894 (1998);
C.\ Z.\ Li, A.\ Bogozi, W.\ Huang, and N.\ J.\ Tao, Nanotech. {\bf 10}, 221 (1999).

\bibitem{ElhoussineAPL02}
F.\ Elhoussine, S.\ M\'at\'efi--Tempfli, A.\ Encinas, and L. Piraux,
Appl. Phys. Lett. {\bf 81}, 1681 (2002).

\bibitem{CostaSS95}  
J.\ L.\  Costa-Kr\"amer, N.\ Garc\'\i a, P.\ Garc\'{\i}a-Mochales, and P.\ A.\ Serena, 
Surf. Sci. {\bf 342}, L1144-L1149 (1995); Erratum in Surf. Sci. {\bf 349}, L138 (1996).

\bibitem{GillinghamJPCM02}
D.\ M.\ Gillingham, I.\ Linington, and J.\ A.\ C.\ Bland,
J. Phys.: Condens. Matter {\bf 14} L567 (2002);
D.\ M.\ Gillingham, C.\ M\"uller, and J.\ A.\ C.\ Bland,
J. App. Phys. {\bf 95} 6995 (2004).

\bibitem{YansonNature99}
A.\ I.\ Yanson, I.\ K.\ Yanson, and J.\ M.\ van Ruitenbeek,
Nature (London) {\bf 400}, 144 (1999).

\bibitem{YansonPRL97}  
A.\ I.\ Yanson and J.\ M.\ van Ruitenbeek, 
Phys. Rev. Lett.  {\bf 79}, 2157 (1997).

\bibitem{HasmyPRL01}
A.\ Hasmy, E.\ Medina, and P.\ A.\ Serena,
Phys. Rev. Lett. {\bf 86}, 5574 (2001).

\bibitem{DiazNanotech01}
M.\ D\'{\i}az, J.L. Costa-Kr\"amer, P.A. Serena, E.\ Medina and A.\ Hasmy,
Nanotechnol. {\bf 12}, 118 (2001).

\bibitem{RubioPRL96}  
G.\ Rubio, N.\ Agra\"{\i}t, and S.\ Viera, 
Phys. Rev. Lett. {\bf 76}, 2302 (1996).

\bibitem{ScheerPRL97}  
E.\ Scheer, P.\ Joyez, D.\ Esteve, C.\ Urbina, and M.\ H.\ Devoret, 
Phys. Rev. Lett. {\bf 78}, 3535 (1997).

\bibitem{ScheerNature98}
E.\ Scheer, N.\ Agra\"{\i}t, J.\ C.\ Cuevas, A.\ Levy-Yeyati, B.\ Ludoph, 
A.\ Mart\'{\i}n-Rodero, G.\ Rubio Bollinger, J.\ M.\ van Ruitenbeek, and C.\ Urbina, 
Nature (London), {\bf 394}, 154 (1998);
J.\ C.\ Cuevas, A.\ Levy Yeyati, and A. Mart\'{\i}n--Rodero, 
Phys. Rev. Lett. {\bf 80}, 1066 (1998).

\bibitem{YansonLTP01}
A.\ I.\ Yanson, J.\ M.\ van Ruitenbeek, and I.\ K.\ Yanson,
Low Temp. Phys. {\bf 27},  807 (2001).

\bibitem{YasudaPRB97}  
H.\ Yasuda and A.\ Sakai, 
Phys. Rev. B {\bf 56}, 1069 (1997);
S.\ K.\ Nielsen, Y.\ Noat, M.\ Brandbyge, R.\ H,\ M.\ Smit, K.\ Hansen,
L.\ Y.\ Chen, A.\ I.\ Yanson, F.\ Besenbacher, and J.\ M.\ van Ruitenbeek,
Phys. Rev. B {\bf 67}, 245411 (2003);
D.\ den Boer, O.\ I.\ Shklyarevskii, and S.\ Speller,
Physica B {\bf 395}, 20 (2007).

\bibitem{LiPRB98}  
C.\ Z.\ Li, H.\ Sha, and N.\ J.\ Tao, Phys. Rev. B {\bf 58}, 6775 (1998).

\bibitem{ShuPRL00}
C.\ Shu, C.\ Z.\ Li, H.\ X.\ He, A.\ Bogozi, J.\ S.\ Bunch, and N.\ J.\ Tao, 
Phys. Rev. Lett. {\bf 84}, 5196 (2000).

\bibitem{YansonPRL01}
A.\ I.\ Yanson, I.\ K.\ Yanson and J.\ M.\ van Ruitenbeek,
Phys. Rev. Lett. {\bf 87}, 216805 (2001).

\bibitem{MedinaPRL03}
E.\ Medina, M.\ D\'\i az, N.\ Le\'{o}n, C.\ Guerrero, A.\ Hasmy, 
P.\ A.\ Serena, and J.\ L.\ Costa--Kr\"{a}mer, 
Phys. Rev. Lett. {\bf 91}, 026802 (2003);
M.\ D\'{\i}az, J.\ L.\ Costa-Kr\"{a}mer, E.\ Medina, A.\ Hasmy and P.\ A.\ Serena, Nanotechnol. {\bf 14}, 113 (2003).

\bibitem{MaresPRB04}
A.\ I.\ Mares, A.\ F.\ Otte, L.\ G. Soukiassian, R.\ H.\ M.\ Smit, and J.\ M-.\ van Ruitenbeek,
Phys. Rev. B. {\bf 70}, 073401 (2004);
A.\ I.\ Mares and J.\ M-.\ van Ruitenbeek,
Phys. Rev. B. {\bf 72}, 205402 (2005);
A.\ I.\ Mares, D.\ F.\ Urban, J.\ B\"urki, H.\ Grabert, C.\ A.\ Stafford, and J.\ M.\ van Ruitenbeek,
Nanotechnol. 18, 265403 (2007).

\bibitem{SirventPRB96}  
C.\ Sirvent, J.\ G.\ Rodrigo, S.\ Vieira, L.\ Jurczyszyn, N.\ Mingo, and F.\ Flores,
Phys. Rev. B {\bf 53}, 16086 (1996).

\bibitem{CostaPRB97}  
J.L. Costa-Kr\"amer, Phys. Rev. B {\bf 55}, R4875 (1997).

\bibitem{OttPRB98} 
F.\ Ott, S.\ Barberna, J.\ G.\ Lunney, J.\ M.\ D.\ Coey, P.\ Berthet, A.\ M.\ de Leon--Guevara, and A.\ Revcolevschi,
Phys. Rev. B {\bf 58}, 4656 (1998).

\bibitem{OshimaAPL98}  
H.\ Oshima and K.\ Miyano, Appl. Phys. Lett. {\bf 73}, 2203 (1998).

\bibitem{OnoAPL99}  
T.\ Ono, Y.\ Ooka, H.\ Miyajima, and Y.\ Otani, 
Appl. Phys. Lett. {\bf 75}, 1622 (1999).

\bibitem{KomoriJPSJ99} 
F.\ Komori and K.\ Nakatsuji, 
J. Phys. Soc. Jap. {\bf 68} 3786 (1999).

\bibitem{GarciaPRL99}  
N.\ Garc\'\i a, M.\ Mu\~noz, and Y.\--W.\ Zhao, 
Phys. Rev. Lett. {\bf 82}, 2923 (1999).

\bibitem{TataraPRL99}  G.\ Tatara, Y.\--W.\ Zhao, M.\ Mu\~noz, and N.\ Garc\'{\i}a, 
Phys. Rev. Lett. {\bf 83}, 2030 (1999).

\bibitem{LudophPRB00} 
B. Ludoph and J.M. van Ruitenbeek, 
Phys. Rev. B {\bf 61}, 2273 (2000).

\bibitem{CostaRMF01}
J.\ L.\ Costa--Kr\"amer, F.\ Briones, and P.\ A.\ Serena, 
Rev. Mex. Fis. {\bf 47 S1}, 69 (2001).

\bibitem{OokaJMMM01}
Y.\ Ooka, T.\ Ono, and H.\ Miyajima
J. Magn. Magn. Mater. {\bf 226-230}, 1848 (2001).

\bibitem{ShimizuJMMM02}
M.\ Shimizu, E.\ Saitoh, H.\ Miyajima, and Y.\ Otani,
J. Magn. Magn. Mater. {\bf 239}, 243 (2002).

\bibitem{ChopraPRB02}
H.\ D.\ Chopra and S.\ Z.\ Hua,
Phys. Rev. B {\bf 66}, 020403(R) (2002).

\bibitem{ViretPRB02}
M.\ Viret, S.\ Berger, M.\ Gabureac, F.\ Ott, D.\ Olligs, I.\ Petej,
J.\ F.\ Gregg, C.\ Fermon, G.\ Francinet, and G.\ Le Goff,
Phys. Rev. B {\bf 66}, 220401(R) (2002).

\bibitem{BakkerPRB02}
D.\ J.\ Bakker, Y.\ Noat, A.\ I.\ Yanson, and J.\ M.\ van Ruitenbeek
Phys. Rev. B {\bf 65} 235416 (2002).

\bibitem{RodriguesPRL03}
V.\ Rodrigues, J.\ Bettini, P.\ C.\ Silva, and D.\ Ugarte
Phys. Rev. Lett. {\bf 91}, 096801 (2003).

\bibitem{ElhoussineJAP03}
F.\ Elhoussine, A.\ Encinas, S.\ M\'at\'efi-Tempfli, and L.\ Piraux,
J. Appl. Phys. {\bf 93}, 8567 (2003).

\bibitem{UntiedtPRB04} 
C.\ Untiedt, D.\ M.\ T.\ Dekker, D.\ Djukic, and J.\ M.\ van Ruitenbeek, 
Phys. Rev. B {\bf 69} 081401(R) (2004).

\bibitem{SekiguchiJMMM04} 
K.\ Sekiguchi, M.\ Shimizu, E.\ Saitoh, and H.\ Miyajima,
J. Magn. Magn. Mat. {\bf 282} 143 (2004).

\bibitem{BrandbygePRB95} 
M.\ Brandbyge, J.\ Schiotz, M.\ R.\ Sorensen,
P. Stoltze, K.\ W.\ Jacobsen, J.\ K.\ Norskov, L.\ Olesen, E.\ Laegsgaard,
I.\ Stensgaard, and F. Besenbacher, 
Phys. Rev. B {\bf 52}, 8499 (1995).

\bibitem{SekiguchiJAP05} 
K.\ Sekiguchi, E.\ Saitoh and H.\ Miyajima,
J. Appl. Phys. {\bf 97}, 10B312 (2005).

\bibitem{DiazJMMM06} 
M.\ D\'{\i}az, J.\ L.\ Costa--Kr\"amer and P.\ A.\ Serena,
J. Magn. Magn. Mat. {\bf 305} 497 (2006).

\bibitem{CalvoIEEE07} 
M.\ R.\ Calvo, M.\ J.\ Caturla, D.\ Jacob, C.\ Untiedt, and J.\ J.\ Palacios,
IEEE Nanotechnology Materials and Devices Conference, Gyeongju (2006).

\bibitem{UntiedtPRL07}
C.\ Untiedt, M.\ J.\ Caturla, M.\ R.\ Calvo,  J.\ J.\ Palacios, R.\ C.\ Segers, and J.\ M.\ van Ruitenbeek,
Phys. Rev. Lett. {\bf 98}, 206801 (2007).

\bibitem{ZabalaPRL98} 
N.\ Zabala, M.\ J.\ Puska, and R.\ M.\ Nieminen,
Phys. Rev. Lett. {\bf 80}, 3336 (1998).

\bibitem{SmogunovSS02}
A.\ Smogunov, A.\ Dal Corso, and E.\ Tosatti
Surf. Sci. {\bf 507-510}, 609 (2002);
A.\ Smogunov, A.\ Dal Corso, and E.\ Tosatti
Surf. Sci. {\bf 566-568}, 390 (2004).

\bibitem{SmogunovPRB06}
A.\ Smogunov, A.\ Dal Corso, and E.\ Tosatti
Phys. Rev. B {\bf 73}, 075418 (2006).

\bibitem{JacobPRB05}
D.\ Jacob, J.\ Fern\'{a}ndez-Rossier, and J.\ J.\ Palacios,
Phys. Rev. B {\bf 71}, 220403(R) (2005).

\bibitem{HasmyPRB05}
A.\ Hasmy, A.\ J.\ P\'erez-Jimenez, J.\ J.\ Palacios, P.\ Garc\'{\i}a-Mochales, J.\ L.\ Costa-Kr\"{a}mer, M.\ D\'{\i}az, E.\ Medina, and P.\ A.\ Serena,
Phys. Rev. B {\bf 72}, 245405 (2005).

\bibitem{GarciaMochalesAPA05}
P.\ Garc\'{\i}a-Mochales, S.\ Pel\'aez, P.\ A.\ Serena, E.\ Medina, and A.\ Hasmy, 
Appl. Phys. A {\bf 81}, 1545 (2005).

\bibitem{PaulyPRB06}
F.\ Pauly, M.\ Dreher, K.\ J.\ Viljas, M.\ H\"{a}fner, J.\ C.\ Cuevas, and P.\ Nielaba, 
Phys. Rev. B {\bf 74}, 235106 (2006).


\bibitem{LandmanSci90}
U.\ Landman, W.\ D.\ Luedtke, N.\ A.\ Burnham, and R.\ J.\ Colton, 
Science {\bf 248}, 454 (1990);
W.\ D.\ Luedtke and  U.\ Landman,
Comp. Mat. Sci. {\bf 1}, 1 (1992).

\bibitem{SuttonJPCM90}
A.\ P.\ Sutton and J.\ B.\ Pethica,
J. Phys. Condens. Matter {\bf 2}, 5317 (1990).

\bibitem{LandmanZPD97}
U.\ Landman, R.\ N.\ Barnett, and W.\ D.\ Luedtke, 
Z. Phys. D {\bf 40}, 282 (1997);
A.\ Nakamura, M.\ Brandbyge, L.\ B.\ Hansen, and K.\ W.\ Jacobsen
Phys. Rev. Lett. {\bf 82}, 1538 (1999);
P.\ Jel\'{\i}nek, R.\ P\'{e}rez, J.\ Ortega, and F.\ Flores,
Phys. Rev. B {\bf 68}, 085403 (2003).

\bibitem{daSilvaPRL01}
E.\ Z.\ da Silva, A.\ J.\ R.\ da Silva, and A. Fazzio,
Phys. Rev. Lett. {\bf 87}, 256102 (2001).

\bibitem{TorresPRL96}  
J.A. Torres and J.J. S\'aenz, 
Phys. Rev. Lett. {\bf 77}, 2245 (1996);
A.\ M.\ Bratkowsky and S.\ N.\ Rashkeev,
Phys. Rev. B {\bf 53}, 13074 (1996);
P.\ Garc\'{\i}a-Mochales and P.\ A.\ Serena,
Phys. Rev. Lett. {\bf 79}, 2316 (1997);
J.\ B\"urki, C.\ A.\ Stafford, X.\ Zotos, and D.\ Baeriswyl,
Phys. Rev. B {\bf 60}, 5000 (1999);
J.\ Zhao, C.\ Buia, J.\ Han and J.\ P.\ Lu, 
Nanotechnology {\bf 14}, 501 (2003).

\bibitem{BratkovskyPRB95}
A.\ M.\ Bratkovsky, A.\ P.\ Sutton, and T.\ N.\ Todorov,
Phys. Rev. B {\bf 52}, 5036 (1995).

\bibitem{SorensenPRB98}
M.\ R.\ Sorensen, M.\ Brandbyge, and K.\ W.\ Jacobsen,
Phys. Rev. B {\bf 57}, 3283 (1998).

\bibitem{IkedaPRL99}
H.\ Ikeda, Y.\ Qi, T.\ Cagin, K.\ Samwer, W.\ L.\ Johnson, and W.\ A.\ Goddard III,
Phys. Rev. Lett. {\bf 82}, 2900 (1999).

\bibitem{BranicioPRB00}
P.\ S.\ Bran\'{\i}cio and J.\-P.\ Rino, 
Phys. Rev. B {\bf 62}, 16950 (2000).

\bibitem{BahnPRL01} 
S.\ R.\ Bahn abd K.\ W.\ Jacobsen,
Phys. Rev. Lett. {\bf 87}, 2661101 (2001).

\bibitem{HeemskerkPRB03}
J.\ W.\ T.\ Heemskerk, Y.\ Noat, D.\ J.\ Bakker, J.\ M.\ van Ruitenbeek, B.\ J.\ Thijsse, and P.\ Klaver, Phys. Rev. B {\bf 67}, 115416 (2003).

\bibitem{WangPhysE05}
B.\ Wang, D.\ Shi, J.\ Jia, G.\ Wang, X.\ Chen and J.\ Zhao, 
Physica E {\bf 30}, 45 (2005).

\bibitem{ParkPRL05}
H.\ S.\ Park, K.\ Gall and J.\ Zimmerman, Phys. Rev. Lett. {\bf 95} 255504 (2005).

\bibitem{DreherPRB05}
M.\ Dreher, F.\ Pauly, J.\ Heurich, J.\ C.\ Cuevas, E.\ Scheer, and P.\ Nielaba, 
Phys. Rev. B {\bf 72}, 075435 (2005).

\bibitem{DawPRL83}
M.\ S.\ Daw and M.\ I.\ Baskes, Phys. Rev. Lett. {\bf 50}, 1285 (1983);
S.\ M.\ Foiles, Phys. Rev. B {\bf 32}, 3409 (1985).

\bibitem{MishinPRB99}
Y.\ Mishin, D.\ Farkas, M.\ J.\ Mehl, and  D.\ A.\ Papaconstantopoulos,
Phys. Rev. B {\bf 59}, 3393 (1999).

\bibitem{PelaezPSS06}
S.\ Pel\'aez, P.\ Garc\'{\i}a-Mochales and P.\ A.\ Serena, 
Physica Stat. Sol. (a) {\bf 203}, 1248 (2006);
S.\ Pel\'aez and P. A. Serena, Surf. Sci. (in press) (2007) (doi:10.1016/j.susc.2007.04.184)

\bibitem{SharvinSPJEPT65} 
Y.\ V.\ Sharvin, Zh. Eksp. Teor. Fiz. {\bf 48}, 984 (1965); Sov. Phys. JEPT 
{\bf 21}, 655 (1965);
J.\ A.\ Torres, J.\ I.\ Pascual, and J.\ J.\ S\'aenz,
Phys. Rev. B {\bf 49}, 16581 (1994).

\bibitem{HerrmannPRL84} H.\ J.\ Herrmann, D.\ C.\ Hong and H.\ E.\ Stanley, Phys. Rev. Lett. {\bf 53}, 1121 (1984); H.\ J.\ Herrmann and H.\ E.\ Stanley, J. Phys. A: Math. Gen. {\bf 17}, L261 (1984).

\bibitem{MehrezPRB97}  
H.\ Mehrez and S.\ Ciraci, 
Phys. Rev. B {\bf 56}, 12632 (1997).

\bibitem{SenPRB02}
P.\ Sen, O.\ G\"ulseren, T.\ Yildirim, I.\ P.\ Batra, and S.\ Ciraci,
Phys. Rev. B {\bf 65}, 235433 (2002).

\bibitem{GonzalezPRL04}
J.\ C.\ Gonz\'alez, V.\ Rodrigues, J.\ Bettini, L.\ G.\ C. Rego, A.\ R.\ Rocha, P.\ Z.\ Coura, S.\ O.\ Dantas, F.\ Sato, and D.\ S.\ Galvao, and D.\ Ugarte,
Phys. Rev. Lett. {\bf 93}, 126103 (2004).

\bibitem{GulserenPRL98}
O.\ G\"ulseren, F.\ Ercolessi and E.\ Tosatti,
Phys. Rev. Lett. {\bf 80}, 3775 (1998).

\bibitem{YansonNature98}
A.\ I.\ Yanson, G.\ Rubio Bollinger, H.\ E.\ van den Brom, N.\ Agra\"{\i}t, J.\ M.\ van Ruitenbeek,
Nature {\bf 395}, 783 (1998);
R.\ H.\ M.\ Smit, C.\ Untiedt, A.\ I.\ Yanson, and J.\ M.\ van Ruitenbeek,
Phys. Rev. Lett. {\bf 87}, 266102 (2001).

\bibitem{SatoAPA05}
F.\ Sato, A.\ S.\ Moreira, P.\ Z.\ Coura, S.\ O.\ Dantas, S.\ B.\ Legoas, D.\ Ugarte, and D.\ S.\ Galvao,
Appl. Phys. A {\bf 81}, 1527 (2005).

\bibitem{RegoPRB03}
L.\ G.\ C.\ Rego, A.\ R.\ Rocha, V.\ Rodrigues and D.\ Ugarte, 
Phys. Rev. B {\bf 67}, 045412 (2003).

\bibitem{PuJCP07}
Q.\ Pu, Y.\ Leng, L.\ Tsetseris, H.\ S.\ Park, S.\ T.\ Pantelides and P.\ T.\ Cummings,
J. Chem. Phys. {\bf 126}, 144707 (2007).

\bibitem{MehrezPRB98}  
H.\ Mehrez and S.\ Ciraci, 
Phys. Rev. B {\bf 58}, 9674 (1998).

\bibitem{LiAPL00}
C.\ Z.\ Li, H.\ X.\ He, A.\ Bogozi, J.\ S.\ Bunch, and N.\ J.\ Tao,
Appl. Phys. Lett. {\bf 76}, 1333 (2000);
Sz.\ Csonka, A.\ Halbritter, G.\ Mih\'{a}ly, E.\ Jurdik, O.\ I.\ Shklyarevskii, S.\ Speller and H.\ van Kempen,
Phys. Rev. Lett. {\bf 90}, 116803 (2003).

\bibitem{SerenaRMF01}
P.\ A.\ Serena and J.\ L.\ Costa-Kr\"amer, 
Rev. Mex. Fis. {\bf 47 S1}, 72 (2001).

\end{thebibliography}
\end{document}